\documentclass[9pt,twocolumn,twoside]{pnas-new}

\makeatletter
\newcommand{\manuallabel}[2]{\def\@currentlabel{#2}\label{#1}}
\makeatother

\setboolean{displaywatermark}{false}
\templatetype{pnasresearcharticle} 
\usepackage{amsfonts}
\usepackage{upgreek}
\usepackage{slashed}
\usepackage{amsmath,amssymb,bm,bbm} 
\usepackage{latexsym}
\usepackage[export]{adjustbox}

\usepackage[colorlinks=true]{hyperref}  
\hypersetup{
    unicode=false,          
    pdftoolbar=true,        
    pdfmenubar=true,        
    pdffitwindow=false,     
    pdfstartview={FitH},    
    pdftitle={Deconfined criticality between an antiferromagnetic insulator and a nodal d-wave superconductor: a quantum Monte Carlo study},    
    pdfauthor={Chuang Chen, Subir Sachdev, Zi Yang Meng},     
    pdfsubject={},   
    pdfcreator={},   
    pdfproducer={}, 
    pdfkeywords={} {} {}, 
    pdfnewwindow=true,      
    colorlinks=true,       
    linkcolor=blue, 
    citecolor=blue,        
    filecolor=magenta,      
    urlcolor=blue           
} 

\usepackage{xcolor}
\definecolor{brandeisblue}{rgb}{0.0, 0.44, 1.0}

\usepackage{mathtools}

\newcommand{\bk}{\mathbf{k}}
\newcommand{\bq}{\mathbf{q}}

\usepackage[toc,page]{appendix}
\usepackage{dsfont}
\usepackage{pdfpages}
\usepackage{graphicx}
\usepackage{subfig}
\usepackage{siunitx}
\pdfoutput=1

\title{Deconfined criticality between an antiferromagnetic insulator and a nodal d-wave superconductor: a quantum Monte Carlo study}

\author[a,b]{Chuang Chen}

\author[c,d]{Subir Sachdev\textsuperscript{1}}

\author[a,b]{Zi Yang Meng}

\affil[a]{Department of Physics and HK Institute of Quantum Science \& Technology, The University of Hong Kong, Pokfulam Road,  Hong Kong SAR, China}
\affil[b]{State Key Laboratory of Optical Quantum Materials, The University of Hong Kong, Pokfulam Road,  Hong Kong SAR, China}

\affil[c]{Department of Physics, Harvard University, Cambridge MA-02138, USA}

\affil[d]{Center for Computational Quantum Physics, Flatiron Institute, 162 5th Avenue, New York, NY 10010, USA}

\leadauthor{Chen} 

\significancestatement{
The hole-doped cuprates are the highest-temperature superconductors known at ambient pressure, yet the origin of their remarkable superconductivity remains unresolved. At low temperatures, these materials can exist either as superconductors or as antiferromagnets, in which neighboring electron spins align in opposite directions. A long-standing idea is that superconductivity is connected to exotic quantum states in which the electron effectively fractionalizes into emergent excitations that carry its spin and charge separately. Using quantum Monte Carlo simulations, we find evidence for such electron fractionalization at the quantum phase transition between antiferromagnetic and superconducting states in a model describing the zero-doping limit. Our results identify a setting in which electron fractionalization can be established numerically, and suggest that it plays a central role in the physics underlying high-temperature superconductivity.
}

\correspondingauthor{\textsuperscript{1}To whom correspondence should be addressed. E-mail: sachdev@g.harvard.edu}


\begin{abstract}
We present a quantum Monte Carlo study of the transition between the insulating N\'eel state and the nodal $d$-wave superconductor on the square lattice at half-filling. We access a regime of frustrated magnetic order without a sign problem using a parton representation of the electron in terms of fermionic spinons and bosonic chargons. Both partons move in a background $\pi$-flux (so the electron experiences no net flux) and are coupled to a quantum fluctuating SU(2) lattice gauge field. In contrast to earlier studies directly on the electronic degrees of freedom, we find evidence for a second-order deconfined quantum phase transition at which both the N\'eel and $d$-wave superconductivity orders vanish continuously. 
We compute correlators of the spinon-chargon composite with the same quantum numbers as the electron: we find 
a gapless Dirac dispersion inside the $d$-wave superconductor, 
turning into a gapped dispersion in the antiferromagnet.
\end{abstract}

\dates{This manuscript was compiled on \today}

\begin{document}

\maketitle
\ifthenelse{\boolean{shortarticle}}{\ifthenelse{\boolean{singlecolumn}}{\abscontentformatted}{\abscontent}}{}



\dropcap{T}here is a well-known connection between the magnetism and superconductivity of metals. This was originally introduced in the context of $^3$He, where ferromagnetic quantum fluctuations are the glue which binds fermionic $^3$He atoms in pairs, leading to varieties of $p$-wave superfluidity \cite{vollhardt2013}. A similar mechanism involving antiferromagnetic spin fluctuations leading to $d$-wave superconductivity appears in the context of the heavy fermion compounds \cite{miyake1986spin,scalapino1986d,bealmonod1986possible}, the pnictides \cite{ChubukovPairing}, and the cuprates \cite{scalapino1995case}. In many of these materials, the onset of antiferromagnetism appears within the `superconducting dome' \cite{Pfleiderer09, ChubukovManifesto}, providing evidence for metallic antiferromagnetic fluctuations ({\it i.e.\/} spin $S=1$ `paramagnons')  as the pairing glue for superconductivity.

Important exceptions to the `antiferromagnetism within the superconducting dome' phase diagram are the hole-doped cuprates, which, likely not co-incidentally, also display the highest superconducting critical temperatures. Here, there is a direct transition at a low doping from long-range antiferromagnetic order to $d$-wave superconductivity, with little or no overlapping region \cite{KeimerHigh}, and this serves as the origin of superconductor-metal-insulator criticality \cite{singlelayer26}. There have long been proposals that the hole-doped cuprates are connected to a parent quantum spin liquid with fractionalized `spinon' and `chargon' excitations in an insulator \cite{LeeNagaosaWen}. More recently, it has been proposed \cite{Christos:2023oru,ChristosLuo24} that hole-doped cuprate superconductivity is connected to a deconfined quantum critical point (DQCP) exhibiting a direct transition, on half-filled single band models on the square lattice, from an antiferromagnetic insulator to a $d$-wave superconductor with nodal, gapless, Bogoliubov quasiparticles (an earlier work \cite{RanVishwanath-easyplane} considered easy-plane antiferromagnets and a fully gapped superconductor, but neither feature is present in the cuprates). The spinons and chargons interact strongly with each other via a SU(2) gauge field at the DQCP, and the low energy state is a conformal field theory with no well-defined quasiparticle excitations. Upon moving away from half-filling, there are corrections which remove the conformal structure at low energies, but the main features of the non-zero doping phase diagram are inherited from the DQCP at half-filling \cite{Christos:2023oru,Sayantan25,Boulder25}.

A number of earlier numerical studies have examined single-band Hubbard-type models on the square lattice at half-filling by direct Monte Carlo evaluation of the path integral of the underlying electrons \cite{AssaadImada,Imada17, Assaad22,Assaad24,Assaad25,Xu:2020qbj,Scaletter21,Scaletter22,HongYao21,HongYao22,HongYao25,Zhaoyu22,Zhaoyu24}, and several of them show proximity between antiferromagnetism and $d$-wave superconductivity. 
Distinct from these works, this paper will investigate the DQCP proposal by carrying out the quantum Monte Carlo (QMC) simulations on a representation that fractionalizes the electrons~\cite{xuMonte2019,chenFermi2021,chenEmergent2025}, and performs the path integral over a resulting SU(2) lattice gauge theory with matter fields associated with
fermionic spinons (which are electrically neutral, but carry spin $S=1/2$) and bosonic chargons (which are spinless but carry electrical charges $\pm e$). 
The benefit of this fractionalized representation is that it allows us to frustrate the antiferromagnetic (N\'eel) order simply by varying a gauge coupling constant. In the earlier studies of Hubbard-type models, this would require second-neighbor exchange interactions, which induces a sign problem. Without such frustration, the N\'eel order remains strong, and the transition to superconductivity is invariably strongly first order. Consequently, our approach
gives us the greater flexibility needed in searching for and studying the vicinity of the DQCP between the antiferromagnetic and $d$-wave superconducting phases.

Underlying the DQCP is an insulating mean-field spin liquid in which free fermionic spinons hop on the square lattice with $\pi$-flux per plaquette \cite{AM88}; the spinons have the dispersion of massless Dirac fermions at two distinct points in the Brillouin zone. Beyond mean-field, these spinons are coupled to an emergent SU(2) gauge field \cite{WenPSG}. Section~\ref{sec:results}.\ref{sec:II} examines spin fluctuations alone by a QMC study of the simplest square lattice gauge action with the spinons minimally coupled to the SU(2) gauge field, along with a Yang-Mills term for the SU(2) gauge field. We present evidence that this model has long-range N\'eel order. So the SU(2) gauge field has confined the massless Dirac spinons, and the N\'eel order is the analog of chiral symmetry breaking in 3+1 dimensional quantum chromodynamics.

We then add charge fluctuations to this insulating antiferromagnet in 
Section~\ref{sec:results}.\ref{sec:chargon} by introducing a bosonic Higgs field (the `chargon') on each site of the square lattice in our QMC simulation. This Higgs field consists of two complex scalars $(B_1, B_2)$, transforms as a fundamental under the gauge SU(2), and also carries a charge of $+e$ under the global U(1) symmetry associated with the electromagnetic field. The chargons must also hop on a background $\pi$-flux, so the physical electron, as a SU(2) gauge-invariant spinon-chargon composite, experiences zero flux. 
The Higgs field Lagrangian can be deduced from symmetry considerations, or derived directly from the Hubbard model or via the ancilla approach \cite{HermeleHoneycomb,Christos:2023oru,Sayantan25,Boulder25}. 
We restrict our attention to the half-filled, particle-hole symmetric case, for otherwise QMC has a sign problem~\cite{panSign2024}; in this case, the bosons have a relativistic dispersion at low energies. Our spectral analysis clearly shows that the composite electron degree of freedom exhibits Dirac dispersion with gapless point at $(\pi/2,\pi/2)$ inside the $d$-wave superconductor (dSC) phase. The corresponding spin dynamic spin structure factor, constructed from the spinon operator, exhibits gapless points at (0,0), $(\pi,0)$ and $(\pi,\pi)$, along with a continuum above these momenta, closely resembling that of the Dirac spin liquid~\cite{maDynamical2018,chenEmergent2025}. As we tune the transition from dSC to AFM phase, the Dirac dispersion of the composite fermions is gapped, and the spin structure in the AFM phase constructed from the spinons develops a clear signal of the spin wave spectrum with a gapless point at $(\pi,\pi)$. At the transition point, enhanced fluctuations in both AFM and valence bond solid (VBS) channels emerge, consistent with the DQCP description \cite{Christos:2023oru,ChristosLuo24}.

\begin{figure}[htp!]
\includegraphics[width=\columnwidth]{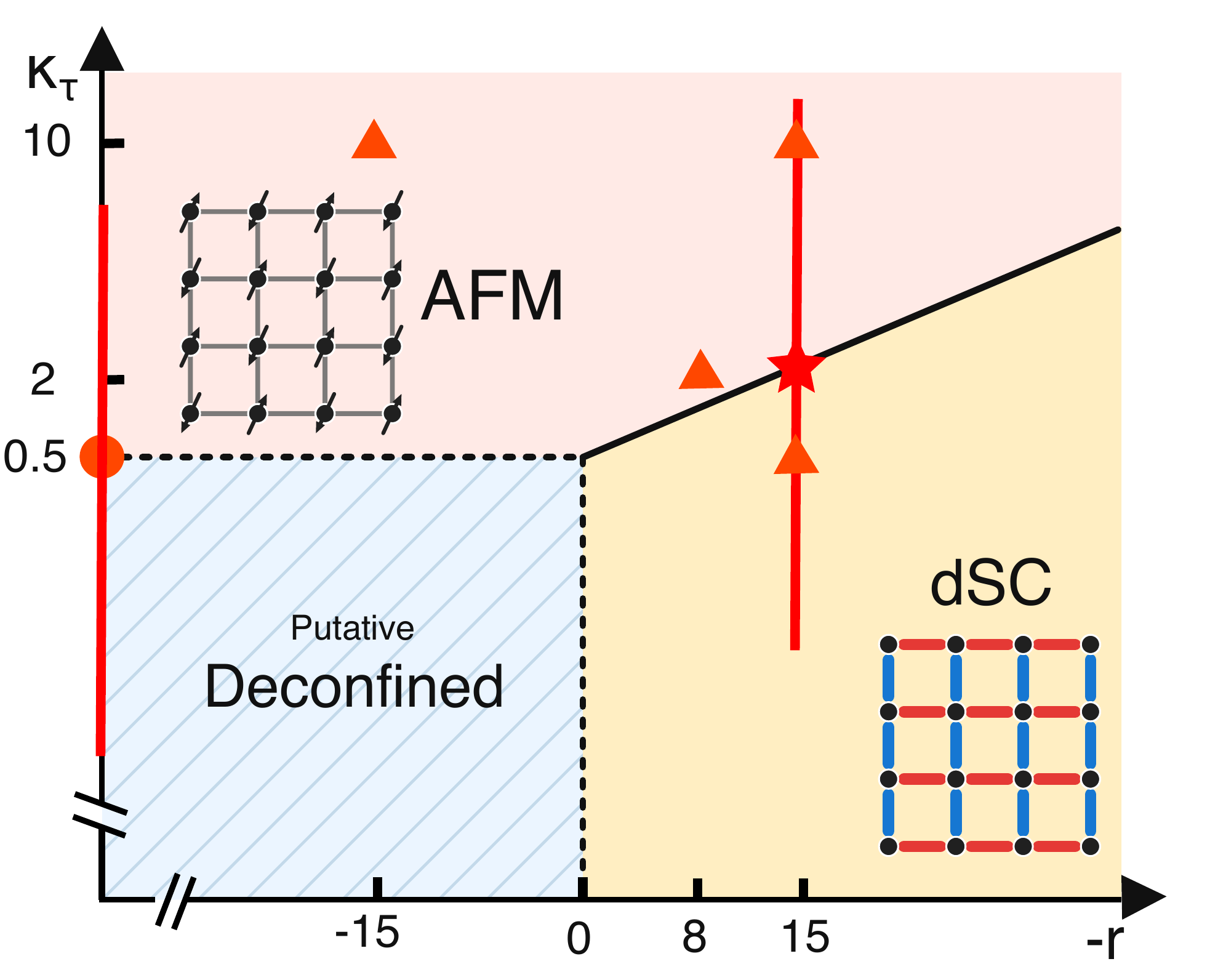}
\caption{\textbf{Phase diagram obtained from QMC simulation.} The $x$-axis is $-r$ and $y$-axis
        is $\kappa_\tau$. The diagram shows the d-wave superconductor
        (dSC), antiferromagnetic Mott insulator (AFM), and a putative
        deconfined regime of the SU(2) gauge theory. The insets schematically illustrate
        the staggered spin configuration of the AFM phase and the opposite
        pairing signs on the horizontal and vertical bonds of the dSC phase. The dots (star and circle) on the
        phase boundaries indicate approximate transition locations inferred
        from the parameter scans in Figs.~\ref{fig:deconfined_to_AFM} and
        \ref{fig:dsc_to_AFM}. The four triangular dots mark the
        parameters of the spectral study (Fig.~\ref{fig:spec}).}
        \label{fig:phase_diagram}
\end{figure}

\section{Results}
\label{sec:results}

\subsection{Spinon matter alone}
\label{sec:II}

We begin by describing the SU(2) lattice gauge theory with fermionic spinons $f_{i \alpha}$, $\alpha = \uparrow, \downarrow $ on the sites, $i$, of the square lattice. This can be derived from the $\pi$-flux saddle point of an insulating antiferromagnet with Heisenberg exchange interactions \cite{AM88,Christos:2023oru}. The Lagrangian can be decomposed as
\begin{equation}
\mathcal{L}_{s} =\mathcal{L}_{f}+\mathcal{L}_{U}+\mathcal{L}_{U_{\tau}}\,,
\label{eq:eq1}
\end{equation}
and the lattice degrees of freedom are the Nambu spinor $\Psi_i$ and the SU(2) gauge field $U_{ij}$ ($i,j$ nearest neighbors)
\begin{align}
\Psi_{i}\equiv\left(\begin{array}{c}
f_{i\uparrow}\\
f_{i\downarrow}^{\dagger}
\end{array}\right), \quad U_{ij}\equiv\left(\begin{array}{cc}
u_{11} & u_{12}\\
-u_{12}^* & u_{11}^*
\end{array}\right), ~\det U_{ij}=1\,.
\end{align}
The spatial part of the fermion Lagrangian is
\begin{align}
\mathcal{L}_{f}  =iJ\sum_{\langle i,j\rangle}e_{ij}(\Psi_{i}^{\dagger}U_{ij}\Psi_{j}-\Psi_{j}^{\dagger}U_{ji}\Psi_{i})
\end{align}
where the background $\pi$-flux is realized by the $e_{ij}$ which we choose
 $e_{ij}=-e_{ji},e_{i,i+\hat{x}}=1,e_{i,i+\hat{y}}=(-1)^{x}$. We set $J=1$ as the energy unit.
 
The dynamics of the SU(2) gauge fields are implemented in spatial
and temporal Yang-Mills terms
\begin{align}
    \mathcal{L}_{U} & =\kappa\sum_{\Box}\left\{
  1-\frac{1}{2}\text{ReTr}\prod_{ij\in\Box}U_{ij}\right\} \nonumber \\
  \mathcal{L}_{U_{\tau}} &
=\frac{1}{\kappa_{\tau}\Delta\tau^{2}}\sum_{ij,\tau}\left\{
  1-\frac{1}{2}\text{Re}\text{Tr}\left[U_{ij}^{\dagger}\left(\tau\right)U_{ij}\left(\tau+1\right)\right]\right\}
\end{align}
The $\kappa$ and $\kappa_\tau$ are
coupling constants and $\Delta\tau$ is the discrete imaginary time step. We use
$\Delta \tau = 0.1$ in QMC simulations as a balance between efficiency and
accuracy.

\begin{figure}[t!]
\includegraphics[width=\columnwidth]{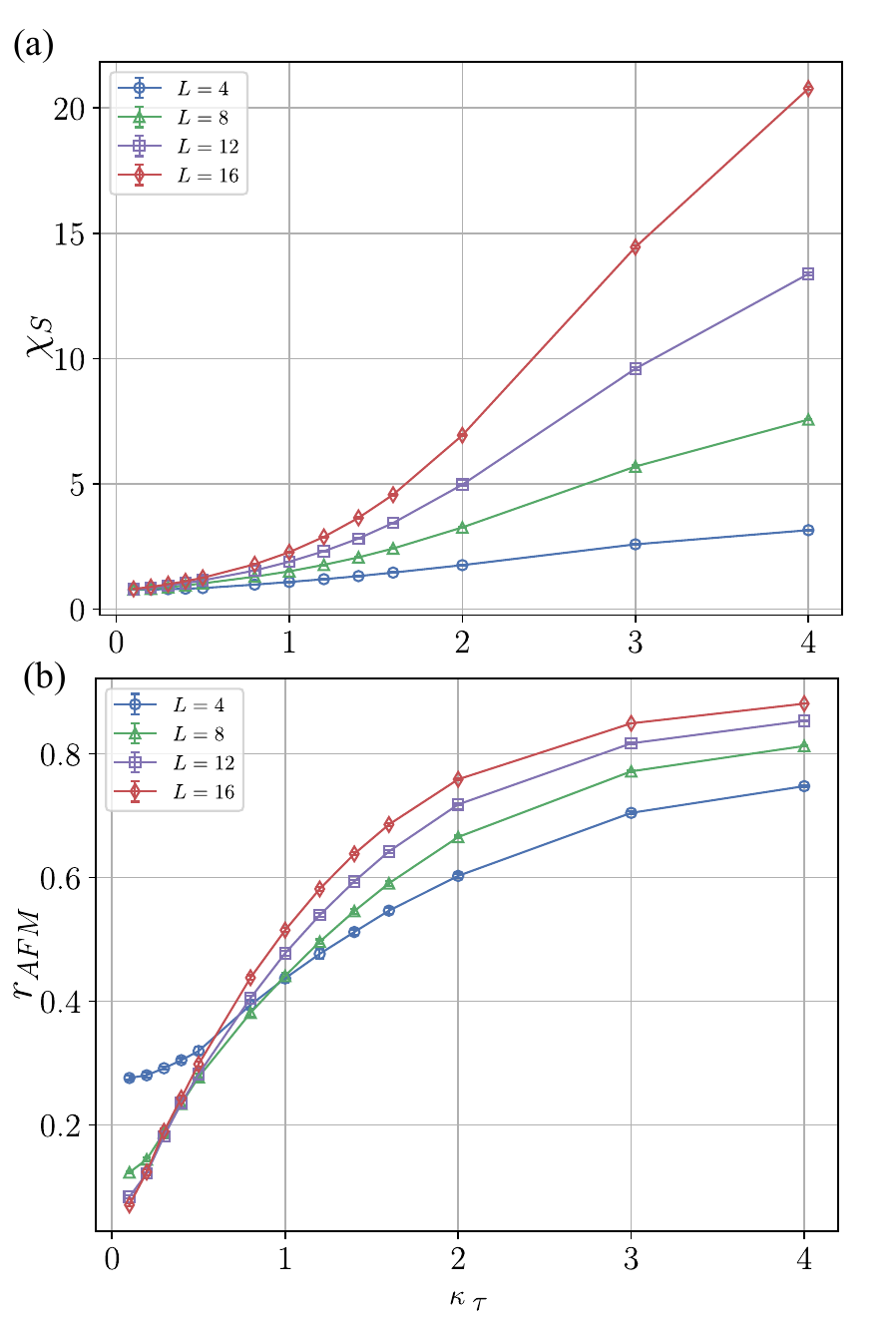}
\caption{\textbf{AFM structure factor and correlation ratio across Deconfined-to-AFM transition}. (a) The structure factor of the AFM
  phase $\chi_{S}$ is measured at the AFM ordered wave vector $\mathbf{q}=\mathbf{M}$ as a function
  of $\kappa_\tau$. Here we fix $J=1$ as the energy unit and $\kappa=20$ and the
  simulations are performed with linear system sizes $L=4,8,12,16$ and inverse
  temperature $\beta=2L$. As $\kappa_\tau$ increases, the AFM order also becomes clearer. (b) The correlation ratio of the AFM
  phase $r_{\text{AFM}}$ as a function
  of $\kappa_\tau$. The crossing of the correlation ratio suggests a putative
quantum critical point between a putative deconfined phase of $\pi$-flux spinons coupled to a fluctuating SU(2) gauge field and the confined AFM phase.}
\label{fig:deconfined_to_AFM}
\end{figure}

To be able to simulate the model in \eqref{eq:eq1}, one needs to perform a
particle-hole transformation for spin-down species of the spinon and rewrite the
fermion determinant, and then one needs to perform the Monte Carlo update of the
SU(2) gauge field $2\times2$ matrices (we introduce a quaternion representation)
on each lattice bond and update SU(2) gauge field $U_{ij}$ with Metropolis fast
update and global update. The computation of the physical observables is performed in each configuration of the SU(2) gauge fields and the bosonic chargon field (as will be mentioned later), with the Wick-decomposition of the spinon Green's function. We leave the detailed description of QMC updates and measurements of observables in the SI~\cite{SI}.

The possible transition between confined and putative deconfined regimes of the SU(2) gauge field can be explored by fixing $\kappa$ and varying $\kappa_{\tau}$. As shown along the left axis $r=+\infty$ (denoted by the left red parameter scan) in the phase diagram of Fig.~\ref{fig:phase_diagram} (corresponds to the no chargon limit). Small $\kappa_\tau$ may correspond to a putative deconfined phase where the Dirac spinons are strongly interacting via fluctuations of the gauge field, while large $\kappa_\tau$ confines the gauge field and the spinons are gapped and develop the N\'eel (AFM) order on the square lattice.

We perform the simulation with system sizes $L=4, 8, 12, 16$ and scale inverse temperatures $\beta$
with $2L$, ensuring targeting zero temperature physics, and we choose $\kappa=20$ and verify $\kappa_\tau$ to detect the transition.
These results are shown in Fig.~\ref{fig:deconfined_to_AFM} (a) and
(b). As $\kappa_\tau$ is increased, $\chi_{S}$, the structure factor of the AFM order, measured as 
\begin{align}
\chi_S(\mathbf{q})=\frac{1}{L^2}\sum_{i,j} S^+_i S^-_j e^{-i(\mathbf{r}_i-\mathbf{r}_j)\cdot \bq},     \label{eq:chiS}
\end{align}
with $S_i^+ =
f_{i\uparrow}^\dagger f_{i\downarrow}, S_i^- = f_{i\downarrow}^\dagger
f_{i\uparrow}$ and at the ordered wave vector $\mathbf{q}=\mathbf{M}=(\pi,\pi)$, is increasing as $\kappa_\tau$ increases and growing as a function of $L$ at each fixed $\kappa_\tau$. This behavior indicates a strong tendency to form the AFM phase in \eqref{eq:eq1} at finite $\kappa_\tau$. However, to locate the possible transition between the putative deconfined phase and the AFM phase, we employ the standard practice in lattice model simulation and monitor whether there is a clear crossing point for the correlation ratio of the AFM order, 
\begin{align}
    r_{\text{AFM}}=1-\frac{\chi_S(\mathbf{M}+\delta \bq)}{\chi_S(\mathbf{M})},
\end{align}
with
$\delta_{\bq}={2\pi}/{L}$ the momentum resolution of Brillouin zone~\cite{pujariInteraction2016,xuMonte2019,chenEmergent2025}. The results are shown
Fig.~\ref{fig:deconfined_to_AFM} (b). One sees an approximate crossing of $r_{\text{AFM}}$ as a function of $\kappa_\tau$ between different system sizes, near $\kappa^c_\tau \sim 0.5$. With the system sizes currently accessible to us, these results suggest a scenario in which the AFM-ordered phase is followed by a disordered phase. For $\kappa_\tau < \kappa^c_\tau$, the data are consistent with the absence of magnetic order and with a $\pi$-flux Dirac phase of spinons coupled to a fluctuating SU(2) gauge field, possibly realizing the putative deconfined phase. However, the present data do not establish the existence of such a phase conclusively: the crossing points drift with system size, and the apparent disordered region is narrow. We therefore cannot rule out that the putative deconfined phase is absent. For $\kappa_\tau > \kappa_\tau^c$, the AFM N\'eel order is developed. We note that similar crossings in the antiferromagnetic correlation ratio have also been seen in the U(1) gauge field coupled to spinons on the square lattice, where a deconfined Dirac spin liquid phase ($\pi$-flux Dirac spinons coupled to the fluctuating U(1) gauge field) and a confined AFM N\'eel phase (with the spinons gapped out) are shown by QMC simulations on both CPU and GPU~\cite{xuMonte2019,chenEmergent2025,fengScalable2026}.

\begin{figure*}[htp!]
\includegraphics[width=\textwidth]{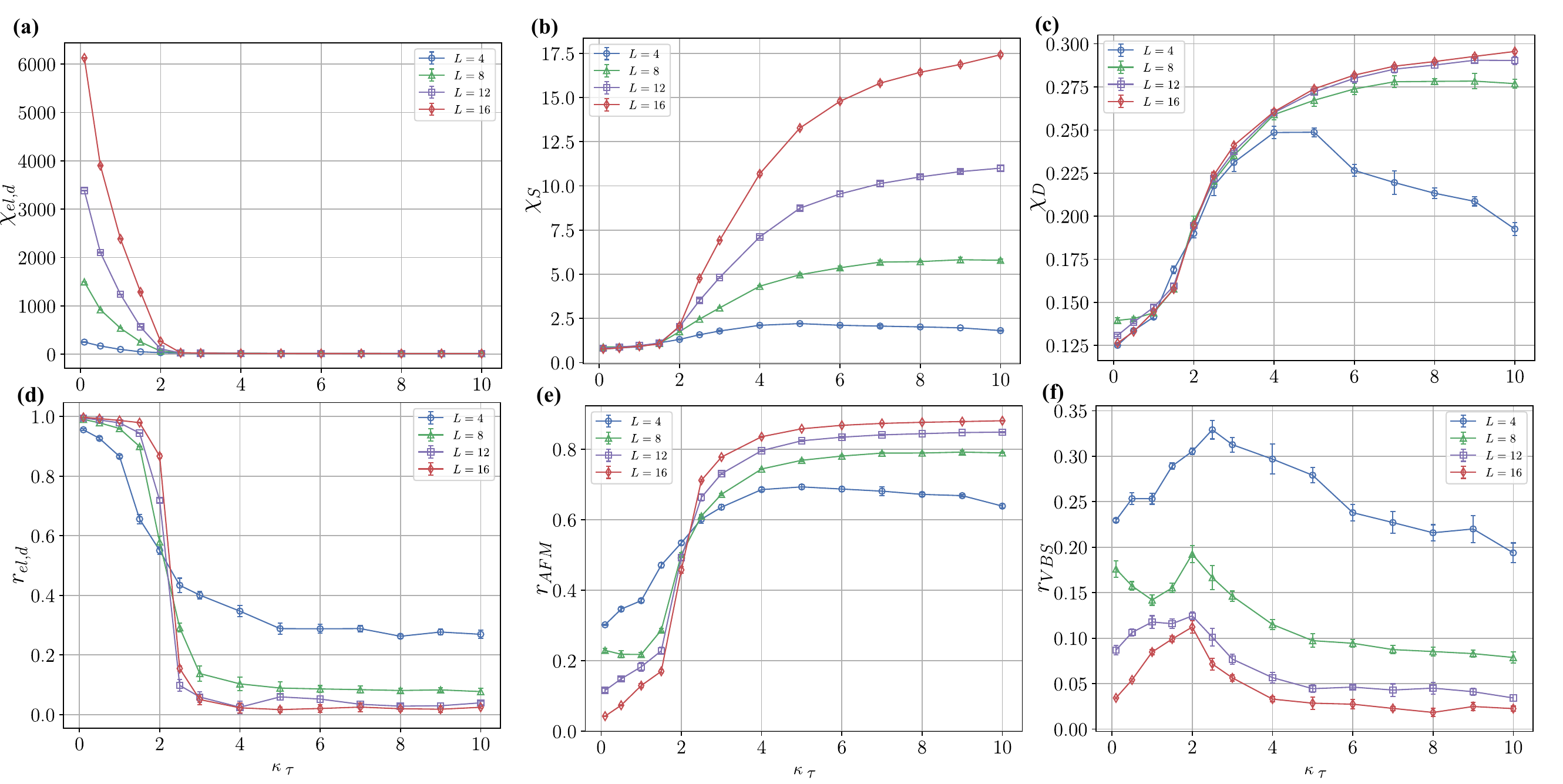}
\centering
\caption{\textbf{Structure factors and correlation ratio of dSC, AFM and VBS
    orders across dSC-to-AFM transition}. We set $\kappa=20, m=0.1, g=0.54/c_0, V_1=4/c_0-3c_0, K_1=3c_0, J_1=3c_0$ in \eqref{eq:eq2}
  and perform the QMC simulations along the path of $r=-15$ and varying $\kappa_\tau$ (the right red scan in Fig.~\ref{fig:phase_diagram}), with linear system sizes $L=4,8,12,16$ and inverse
  temperature $\beta=2L$. (a) The structure factor of the dSC
  phase $\chi_{\text{el, d}}$ as a function
  of $\kappa_\tau$. As $\kappa_\tau$ increases, dSC order decreases and
  vanishes. (b) The structure factor of the AFM phase $\chi_S$. As
  $\kappa_\tau$ increases, AFM order starts to increase at the point 
  where dSC order starts to decrease. (c) The structure factor of VBS
  $\chi_\text{D}$. As $\kappa_\tau$ increases, the VBS order parameter
  is mildly growing but not divergent with system sizes, indicating no long-range VBS
  order. (d) The correlation ratio of the dSC
  phase $r_{\text{el, d}}$ as a function of $\kappa_\tau$. Lines for different
  system sizes cross near $\kappa_{\tau, c, \text{dSC}} \sim 2.2(2)$. (e) The correlation
  ratio of the AFM phase $r_\text{AFM}$. The quantum critical point $\kappa_{\tau,c,\text{AFM}}$ is located at $2.2(2)$, whose value is consistent with dSC quantum critical point
  $\kappa_{\tau, c, \text{dSC}}$. (f) VBS correlation ratio $r_\text{VBS}$. Near
the dSC-AFM quantum critical point, VBS correlation ratio of different sizes shows a peak, signifying the enhanced symmetry at the DQCP. }
        \label{fig:dsc_to_AFM}
\end{figure*}

The presence of AF order in the spinon-only lattice gauge theory in Eq.~(\ref{eq:eq1}) has implications for its  continuum description in terms of 2 flavors of Dirac fermions coupled to a SU(2) gauge field \cite{Wang17,ChristosLuo24}: 
\begin{equation}
\mathcal{L}_\psi = i \bar{\psi} \gamma^\mu \left( \partial_\mu - i A^\alpha_{\mu}\sigma^\alpha \right) \psi + \mathcal{L}_4 \label{LF}
\end{equation}
where $\psi$ are $N_f = 2$ Dirac fermions obtained from the continuum limit of $\Psi$ at two Dirac points in the Brillouin zone, 
$A_\mu^\alpha$ is the SU(2) gauge field obtained from the continuum limit of $U_{ij}$, $\gamma^\mu$ are $2\times 2$ Dirac matrices which act on the sublattice space, and
$\sigma^\alpha$ are Pauli matrices.
Wang~{\it et al.\/} \cite{Wang17} argued that a relevant 4-Fermi term in $\mathcal{L}_4$ drove the N\'eel-VBS transition, and so the effective theory of our lattice model must have such a term favoring the N\'eel state.

\subsection{Spinon and chargon matter}
\label{sec:chargon}

Now we add charge fluctuations to the spinon-only insulating model of Section~\ref{sec:results}.\ref{sec:II}, and study the onset of $d$-wave superconductivity (dSC). The charge fluctuations are realized by a bosonic Higgs
field $B_i$ -- the chargon -- on each lattice sites~\cite{Christos:2023oru,Sayantan25,Boulder25}. The Lagrangian for $B_i$ is obtained by symmetry considerations, but can also be derived explicitly from the Hubbard model following the methods of Hermele for the honeycomb lattice \cite{HermeleHoneycomb}.
The overall Lagrangian is now
\begin{equation}
  \mathcal{L} = \mathcal{L}_{s} + \mathcal{L}_{c},
  \label{eq:eq2}
\end{equation}
where the chargon Lagrangian is
\begin{equation}
\mathcal{L}_{c}=\mathcal{E}_2(U,B) + \mathcal{E}_4(U,
B)+\mathcal{L}_{B_{\tau}}.
\end{equation} 
More specifically, 
\begin{align}
\mathcal{E}_2(U,B) &=
(r+2\sqrt{2}w)\sum_i B_i^\dagger B_i \nonumber \\
&+ iw \sum_{\langle ij\rangle} e_{ij} \left(B_i^\dagger U_{ij} B_j - B_j^\dagger U_{ji} B_i \right)
\end{align}
is the gauge-invariant quadratic term, 
\begin{align}
\mathcal{E}_4(U,B)&=\frac{u}{2}\sum_i \rho_i^2 + V_1 \sum_i \rho_i \left(
 \rho_{i+\hat{x}}+\rho_{i+\hat{y}} \right) \nonumber \\ 
 & + g\sum_{\langle ij\rangle}
 |\Delta_{ij}|^2 + J_1 \sum_{\langle ij\rangle} Q_{ij}^2 +
K_1 \sum_{\langle ij\rangle} J_{ij}^2
\end{align}
is the gauge-invariant quartic term, and 
\begin{equation}
\mathcal{L}_{B_{\tau}}=\sum_i \frac1m |D_\tau B_i(\tau)|^2=\sum_i \frac{\left| B_i(\tau+\Delta\tau) -
    U_{i,\tau}(\tau)\,B_i(\tau) \right|^2 }{ m(\Delta\tau)^2}
\end{equation}
is the imaginary
time covariant derivative term (stands for the kinetic energy term), with temporal
gauge link variable $U_{i,\tau}= \mathbf{1}_2$ by our gauge choice. The QMC updates of both SU(2) gauge field and the chargon boson field are given in SI~\cite{SI}.
Ref.~\cite{Christos:2023oru} demonstrated that various gauge-invariant physical observables could be identified: 
\begin{align}
\left\langle c_{i\alpha}^\dagger c_{i\alpha} \right\rangle & \sim \rho_i = B_i^\dagger B_i \nonumber  \\
\left\langle c_{i\alpha}^\dagger c_{j\alpha} + c_{j\alpha}^\dagger c_{i\alpha} \right\rangle
  &  \sim Q_{ij} = Q_{ji} = \operatorname{Im} \left( B_i^\dagger e_{ij} U_{ij} B_j \right) \nonumber  \\
i\left\langle c_{i\alpha}^\dagger c_{j\alpha} - c_{j\alpha}^\dagger c_{i\alpha} \right\rangle & \sim J_{ij}
= - J_{ji} = \operatorname{Re} \left( B_i^\dagger e_{ij} U_{ij} B_j \right) \nonumber  \\
\left\langle \upvarepsilon_{\alpha\beta} c_{i\alpha} c_{j\beta} \right\rangle & \sim \Delta_{ij} = \Delta_{ji}
= \upvarepsilon_{ab} e_{ij} B_{ai} [U_{ij}]_{bc} B_{cj}
\label{eq:gauge_invarants}
\end{align}
with $\rho_i$ the site charge density, $Q_{ij}$ the bond density, $J_{ij}$ the bond
current and $\Delta_{ij}$ the pairing operator, and
\begin{align}
    \upvarepsilon = \left(\begin{array}{cc}
0 & 1\\
-1 & 0
\end{array}\right)
\end{align}
is the anti-symmetric matrix.

By varying $r$ in $\mathcal{E}_2(U,B)$, we can achieve a Higgs transition that
condenses chargon $B$ at $r=0$ and enter various spontaneous symmetry breaking phases.
Following the literatures~\cite{Sayantan25,Boulder25}, we set the appropriate parameters in $\mathcal{E}_4(U,B)$ such as $g=0.54/c_0,
J_1=3c_0, K_1=3c_0$, with $c_0=4(1+\sqrt{2})^2$, we can put the Higgs
phase in an ordered phase that corresponds to $d$-wave superconductor (dSC) of
gauge-invariant electron in the model. Combined with the possible confinement transition by
tuning $\kappa_\tau$ in the spinon Hamiltonian (\eqref{eq:eq1}), the overall phase diagram of Hamiltonian \eqref{eq:eq2} spanned by the axes of 
$\kappa_\tau$ vs $-r$ is shown in Fig.~\ref{fig:phase_diagram}. Note that the
$x$-axis is $-r$ such that the bottom-left corner corresponds to the putative
deconfined regime. The
important property of the phase diagram is that by adding chargon to the spinon
Hamiltonian \eqref{eq:eq1}. We can have an extra dSC phase that transitions
either from the putative deconfined regime or from the AFM phase. And notably,
the transition from AFM to dSC could occur at finite $r$ beyond the possible
transition between the putative deconfined regime and the AFM phase in \eqref{eq:eq1}. Moreover, according to our
QMC data, such a direct dSC-to-AFM transition, seem to be consistent with the DQCP scenario, as we now turn to.

In order to demonstrate the existence of dSC order and determine its transition to the AFM phase, we measure the pairing structure factor
\begin{align}
    \chi_{\text{dSC}}(\bq)=\frac{1}{L^2}\sum_{i,j}\Delta_i^\dagger \Delta_j e^{-i(\mathbf{r}_i-\mathbf{r}_j)\cdot \bq}
\end{align} with
$\Delta_i =\sum_j(-1)^{\eta(j)}\Delta_{ij}$ with $\eta(j)$ is the d-wave form
factor $\{1,-1,1,-1\}$ of the four nearest-neighbor bonds originated from $i$ to $j$ and $\Delta_{ij}$ from \eqref{eq:gauge_invarants}. As the ordering wave vector of dSC is at $\mathbf{q}=\mathbf{\Gamma}$, the structure factor of the dSC order $\chi_{\text{el,d}}=\chi_{\text{dSC}}(\mathbf{\Gamma})$ and the corresponding correlation ratio
\begin{align}
    r_{\text{el, d}}=1-\frac{\chi_{\text{dSC}}(\mathbf{\Gamma}+\delta
  \bq)}{\chi_{\text{dSC}}(\mathbf{\Gamma})}.
\end{align}
The AFM structure factor $\chi_S$ is
defined in the previous section. In order to study the existence of emergent higher symmetry at the DQCP, we measure the valence bond solid (VBS) order parameter from dimer correlation function. We construct $x$-direction dimer operator $D_i = S_{i}
\cdot S_{i+\hat{x}}$ and measure its structure factor \begin{align}
    \chi_{\text{VBS}}(\bq)
=\frac{1}{L^2}\sum_{i,j}D_i D_j e^{-i(\mathbf{r}_i-\mathbf{r}_j)\cdot \bq}
\end{align}
(the detailed QMC implementation of this observable is given in SI~\cite{SI}). The ordering vector is at $\mathbf{X}=(\pi, 0)$ and the corresponding correlation ratio
\begin{align}
    r_{\text{VBS}}=1-\frac{\chi_{\text{VBS}}(\mathbf{X}+\delta\bq)}{\chi_{\text{VBS}}(\mathbf{X})}.
\end{align}

To demonstrate the dSC-to-AFM transition, we fix $r=-15$ with $\kappa=20, m=0.1$
and vary $\kappa_\tau$. This is the right red parameter scan in the phase
diagram in Fig.~\ref{fig:phase_diagram}. The results are shown in
Fig.~\ref{fig:dsc_to_AFM}. From panels (a) and (b), one sees as a function of
$\kappa_\tau$, the dSC order decreases and the AFM order increases, there seems
to be a continuous transition from dSC to AFM around $\kappa_\tau \sim 2$. To
precisely determine the transition point, we again plot the corresponding
correlation ratio of the two orders, the results are shown in
Fig.~\ref{fig:dsc_to_AFM} (d) and (e), where one clearly sees the crossing of
both ratios, consistently located at $\kappa^c_\tau = 2.2(2)$. Such a continuous
transition fits well into the fermionic DQCP scenarios proposed in
Refs.~\cite{Christos:2023oru,ChristosLuo24,Sayantan25,Boulder25}. 

Moreover,
Fig.~\ref{fig:dsc_to_AFM} (c) shows that along the same path, there is no VBS
long-range order in both dSC and AFM phases (as the extensive structure factor
is not divergent with respect to the system sizes and $\chi_{\mathrm{VBS}}/L^2$ will go to zero upon extrapolating to $L\to\infty$). 
As shown in Fig.~\ref{fig:dsc_to_AFM} (f), the correlation ratio $r_{\text{VBS}}$ develops a peak at $\kappa^c_\tau$, signifying the enhanced fluctuations at the DQCP with emergent higher symmetry~\cite{Wang17, gazitConfinement2018}. 
We attribute the monotonic increase in $\chi_{\mathrm{VBS}}$ across the transition to a smooth background evolution.
In the SI~\cite{SI}, we also examine charge density wave (CDW) order near the DQCP, but did not observe any enhancement. This can be understood from the absence of an enhanced symmetry between dSC and CDW; the enhanced symmetry is present at quadratic order, but is broken by relevant quartic terms $\bar{v}_{1,2}$ in the chargon action in Eq.~(\ref{LB}) below.

The continuum description of the AF-dSC DQCP in Ref.~\cite{ChristosLuo24} has the Lagrangian
\begin{equation}
\mathcal{L}_{\rm DQCP} = \mathcal{L}_\psi + \mathcal{L}_B
\end{equation}
where $\mathcal{L}_\psi$ is the fermion component in Eq.~(\ref{LF}). The chargon dispersion has two minima in the Brillouin zone, and so the continuum limit yields a Higgs field $B_{as}$ where $a$ is the fundamental SU(2) gauge index, and $s$ is the valley index taking 2 values; the chargon Lagrangian is
\begin{align}
\mathcal{L}_B = & |(\partial_\mu  - i A^\alpha_\mu \sigma^\alpha) B|^2 + \bar{r} |B|^2 + \bar{u} \left( |B|^2 \right)^2 \nonumber \\
&+ \bar{v}_1 \left| \upvarepsilon_{ab} \upvarepsilon_{st} B_{as} B_{bt} \right|^2   \label{LB} \\
&+ \bar{v}_2 \left[ \left( B_{a1}^\ast B_{a1} - B_{a2}^\ast B_{a2}\right)^2 +  \left( B_{a1}^\ast B_{a2} + B_{a2}^\ast B_{a1} \right)^2 \right] \,.  \nonumber
\end{align}
Here $\bar{r}$ is the tuning parameter across the DQCP, $\upvarepsilon_{ab} \upvarepsilon_{st} B_{as} B_{bt}$ is the continuum limit of the gauge-invariant dSC order parameter, while $B_{a1}^\ast B_{a1} - B_{a2}^\ast B_{a2}$ and $B_{a1}^\ast B_{a2} + B_{a2}^\ast B_{a1}$ 
are the CDW order parameters in the $x$ and $y$ directions.
Our results here indicate that the 4-Fermi term $\mathcal{L}_4$ in Eq.~(\ref{LF}) is {\it dangerously irrelevant} {\it i.e.} it is irrelevant at the DQCP, but relevant on the AF side, once the Higgs field $B$ has been gapped out.

\begin{figure*}[htp!]
  \includegraphics[width=0.95\textwidth]{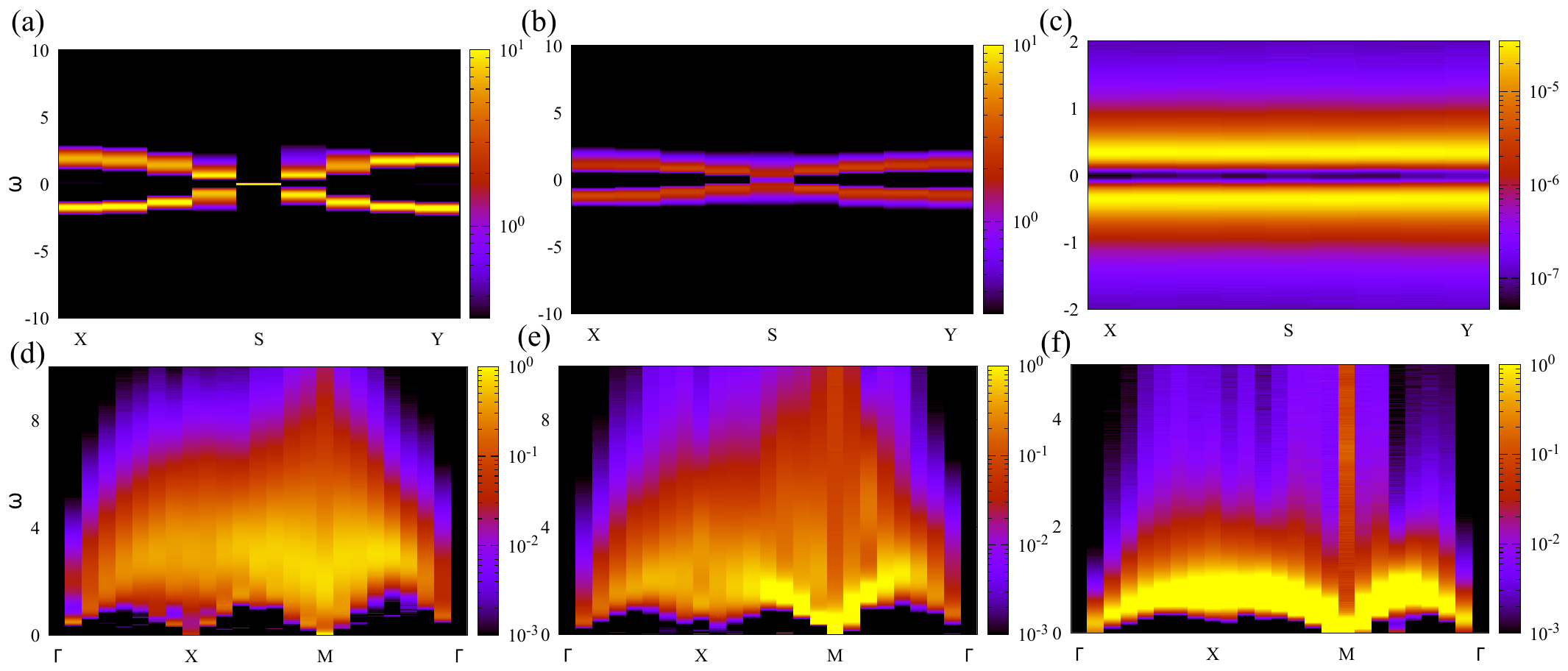}
  \centering
  \caption{\textbf{Electron and spin spectra across dSC-to-AFM transition}. (a) shows the electron spectra
    $A(\bk, \omega)$ inside the dSC phase with $r=-15, \kappa_\tau=1$, while (b) $r=-8,
    \kappa_\tau=2$ and (c) $r=15, \kappa_\tau=10$ are electron spectra inside the AFM phase, with (b) closer to the transition point.
    (d), (e) and (f) are the corresponding spin spectra $\chi_S(\bq,
    \omega)$. For electron spectra $A(\bk, \omega)$,
    we plot the high symmetry path connecting $\mathbf{X}(\pi, 0) \rightarrow \mathbf{S}(\pi/2,
    \pi/2) \rightarrow \mathbf{Y}(0, \pi)$. While for
  the spin spectra, we use path $\mathbf{\Gamma}(0,0) \rightarrow \mathbf{X}(\pi,
  0) \rightarrow \mathbf{M}(\pi, \pi) \rightarrow \mathbf{\Gamma}(0,0)$. In the dSC phase,
  the electron shows Dirac-like dispersion, and the spin spectrum is in the shape of two-spinon continuum of U(1) Dirac spin liquid. In the AFM phase,
  however, close to quantum critical point, the electron spectra slightly opens gap at
  Dirac point and the corresponding spin spectra shows signs of spin-wave-like
  dispersion near $\mathbf{M}$ point and is weakly gapped at $\mathbf{X}$, and the continuum due to DQCP fluctuations are proliferated. While deep in the AFM
  phase, the electron spectra are fully gapped and nearly flat. The
  corresponding spin spectra show spin-wave-like dispersion with Goldstone mode
  at ordering vector $\mathbf{M}$.}
  \label{fig:spec}
\end{figure*}
\subsection{Electron spectral function.} Once we have verified the transition from dSC to AFM, we can study the composite gauge-invariant
fermionic object $c$ dubbed as ``electron'', whose quantum number is equal to
that of a real electron,
\begin{align}
  c_{i\alpha} \sim & B_{1i}^*\, f_{i\alpha} + B_{2i}^*\, \varepsilon_{\alpha\beta} f_{i\beta}^\dagger \nonumber \\
  c_{i\alpha}^{\dagger} \sim & B_{1i}\, f_{i\alpha}^{\dagger} + B_{2i}\, \varepsilon_{\alpha\beta} f_{i\beta}
\end{align}
by computing its dynamic Green's function
\begin{equation}
    G_{c}(\mathbf{k},\tau) = \frac{1}{L^2}\sum_{i,j}\langle c_{i}(\tau) c^\dagger_{j}(0)\rangle e^{-i\mathbf{k}\cdot(\mathbf{r}_i-\mathbf{r}_j)}
\end{equation}
from which the spectral function $A(\mathbf{k},\omega)$ can be obtained from the
stochastic analytic continuation (SAC)~\cite{beachIdentify2004,chenEmergent2025,shaoNearly2017,yang2025spinonsspinchargeseparationdeconfined,shao2023progress}. At the same time, we
can as well measure the dynamical spin structure factor $\chi^\pm_S(\bq, \tau)$ of the spinons in Eq.~(\ref{eq:chiS}), 
and use SAC to obtain the spin spectra $\chi_S(\bq, \omega)$.

To study the spectral properties of the model in dSC and AFM phases, we
choose three parameter points with two deep inside either phases and one close to the transition point. Specifically, for dSC we
choose $r=-15, \kappa_\tau=1$, for AFM we choose $r=15, \kappa_\tau=10$, and we choose $r=-8, \kappa_\tau=2$ close to the transition, as labeled by the three red triangles in Fig.~\ref{fig:phase_diagram}. We measure
both dynamical structure factors $G_c(\bk, \tau)$ and $\chi_S(\bq, \tau)$ at system
size $L=16$ and inverse temperature $\beta=16$.

The resulting electron spectra $A(\bk, \omega)$ as well as spin spectra
$\chi_S(\bq, \omega)$ are shown in Fig.~\ref{fig:spec}. Inside the dSC phase, the electron spectra
shows Dirac-like dispersion, with gapless Dirac point at momentum $\mathbf{S}= (\pi/2, \pi/2)$ (Fig.~\ref{fig:spec} (a)).
Deep in the AFM phase, the electron spectra is fully gapped and nearly flat (Fig.~\ref{fig:spec} (c)). The gap
size is around $\omega \sim 0.5$ and the intensity of the spectra is quite weak
($A(\bk, \omega) \sim 10^{-5}$), which is due to the fact that with positive
$r$, $\langle |B| \rangle \sim 0$, resulting in vanishingly small $G_c(\bk,
\tau)$ and $A(\bk, \omega)$. And close to the transition, as shown in Fig.~\ref{fig:spec} (b),  the spectral weight at the Dirac point is vanishing, and yet the continuum due to the DQCP fluctuations is stronger than that inside the dSC phase.

As for the spin spectra, inside the dSC phase, as shown in Fig.~\ref{fig:spec} (d), it forms the Dirac cone continuum shape~\cite{zengSpectral2024} with the minima at momenta of $\mathbf{X}=(\pi,0)$ and $\mathbf{M}=(\pi,\pi)$, which comes from the convolution of the Dirac dispersion of the electron Green's function. We note, similar U(1) Dirac spin liquid spectra have been seen inside the U(1) deconfined phase of Dirac fermions coupled to the U(1) gauge field~\cite{chenEmergent2025}, and in a few spin models with enhanced U(1) Dirac spin liquid behavior such as that close to the DQCP of the JQ model~\cite{maDynamical2018} and Dirac spin liquid state of $J_1$-$J_2$ triangular lattice Heisenberg model~\cite{ferrariDynamical2019}.
Deep in the AFM phase, as shown in Fig.~\ref{fig:spec} (f), the spin spectra show typical spin-wave-like dispersion, with
$\mathbf{X}$ gapped and $\mathbf{M}$ being gapless Goldstone mode of the N\'eel order~\cite{shaoNearly2017}. However, we also note that even within the AFM phase, the spin-wave spectra remain significantly broadened compared with those of the square-lattice antiferromagnetic model. We believe that such spectral broadening also arises from residual gauge fluctuations in our model, as given by \eqref{eq:eq2}. Close to the dSC-to-AFM transition, the spin spectra in Fig.~\ref{fig:spec} (e) clearly exhibit DQCP behavior, with continua at both the $\mathbf{X}$ and $\mathbf{M}$ points, with $\mathbf{X}$ slightly gapped and the spin-wave at $\mathbf{M}$ starting to manifest, as our parameter is, in fact, slightly located on the AFM side of the transition as shown in Fig.~\ref{fig:phase_diagram}.
In SI~\cite{SI}, we have also added electron and spin spectra deep inside the AFM phase with another set of parameters.

\section{Discussion}

\begin{figure}[htb!]
\includegraphics[width=\columnwidth]{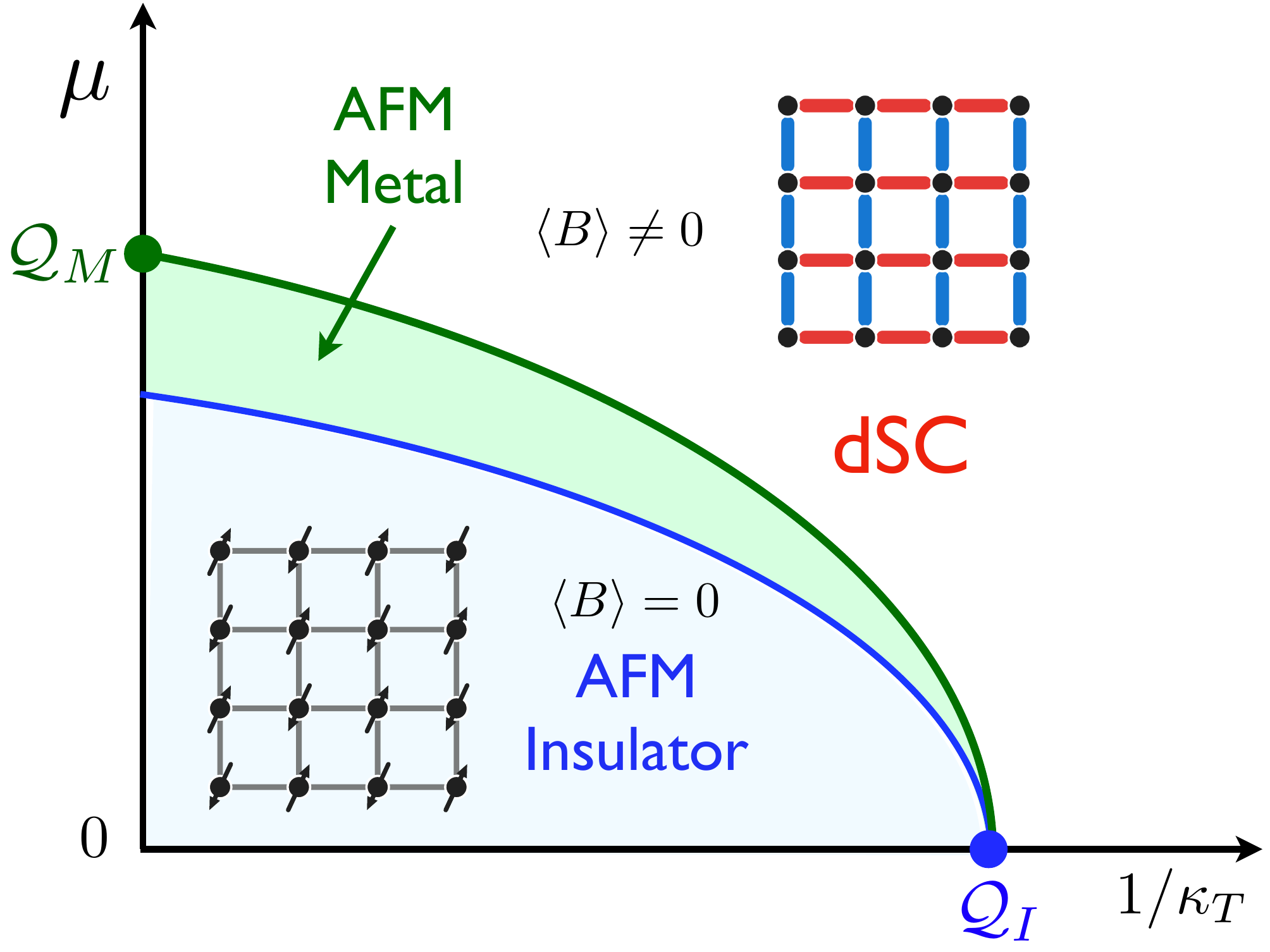}
\caption{\textbf{Proposed $T=0$ phase diagram away from half-filling.} The density is at half-filling when the chemical potential $\mu=0$, and within the shaded AFM Insulator region. The point $\mathcal{Q}_I$ is the DQCP studied in the present paper. The AFM Metal has long-range AFM order and Fermi surfaces of hole pockets each enclosing fractional area $p/4$. The point $\mathcal{Q}_M$ is the $T=0$ AFM Metal-dSC phase transition present in Fig.~2 of Ref.~\cite{KeimerHigh}, which we propose here controls the $T>0$ pseudogap region. The AFM-dSC transition at non-zero $p$ can be weakly first-order or involve co-existence.}
        \label{fig:doping}
\end{figure}

We have presented significant evidence for a DQCP on the half-filled square lattice between the AFM and dSC states. We also showed that the nodal Bogoliubov quasiparticles of the dSC acquire a gap in the AFM state. 

These results provide support for the framework for the hole-doped cuprates described in Refs.~\cite{Christos:2023oru,ChristosLuo24,Sayantan25,Boulder25}.
In these works, such a half-filling DQCP provides a starting point for understanding the quantum states at low doping $p$. This is illustrated  Fig.~\ref{fig:doping}, which extends the half-filling phase diagram to include a non-zero chemical potential $\mu$ at $T=0$.  The DQCP at $\mathcal{Q}_I$ is connected to a nearby quantum phase transition between the AFM Metal and dSC at $\mathcal{Q}_M$. As we noted in the introduction, the phase diagram of the hole-doped cuprates \cite{KeimerHigh} has a direct $T=0$ transition from the AFM Metal to the dSC with increasing doping $p$, and this shows interesting supercondutor-metal-insulator quantum criticality in the single-layer limit \cite{singlelayer26}. The AFM metal has Fermi pockets, and these will induce a relevant linear time-derivative term in the action for the Higgs bosons $B$ at $\mathcal{Q}_M$ \cite{Christos:2023oru}, which will remove the conformal invariance of the quantum phase transition at $\mathcal{Q}_I$. However, the small value of $p$ makes the strength of this perturbation weak, and we expect the basic features of the DQCP to survive at $\mathcal{Q}_M$.

The $T>0$ pseudogap metal at non-zero $p$ is proposed to display the physics of the critical quantum spin liquid liquid underlying the transition at $\mathcal{Q}_M$, and hence at $\mathcal{Q}_I$. There is no symmetry-breaking in the pseudogap, and the quantum spin liquid makes possible a `fractionalized Fermi liquid (FL*)' metal whose Fermi surface does not enclose the Luttinger area \cite{TSSSMV03}. The FL* pseudogap has an additional band of charge $e$, spin $S=1/2$ quasiparticles occupying 4 Fermi pockets, each of fractional area $p/8$, along with the spinon and chargon degrees of freedom considered in the present paper.
This description of the pseudogap is supported by recent angle-dependent magnetoresistance observations \cite{Ramshaw22,Yamaji24,Zhao_Yamaji_25,Fuchun_Yamaji}.

We finally address an important question concerning the implementation of
Gauss's law in numerical simulations of lattice gauge theories. In previous studies of
$\mathbb{Z}_2$ gauge-matter systems, the Gauss-law constraint was either
explicitly imposed by projecting onto a fixed gauge-charge sector or allowed
to emerge dynamically in the low-temperature limit
\cite{gazitConfinement2018,frankEmergence2020}. Similarly, sign-problem-free simulations of compact
$U(1)$ gauge theory coupled to fermionic matter were formulated without an
explicit Gauss-law projection, and the authors argued that the constraint is
dynamically generated in the zero-temperature and thermodynamic limits
\cite{xuMonte2019}. Our present $SU(2)$ simulations are likewise performed in an
unconstrained temporal-gauge formulation, without an explicit integration over
the temporal gauge field. It would therefore be important in future work to
compare this formulation with one in which Gauss's law is explicitly imposed,
for example through a projection onto the physical Hilbert space or by
integrating over the temporal gauge field. In particular, such a comparison
would clarify whether the putative deconfined phase, as well as its
finite-size and infrared properties, is unaffected by the choice of
constraint implementation.



\subsection*{Acknowledgments}
C.C. and Z.Y.M. acknowledge the support from the Research Grants Council (RGC) of Hong Kong Special Administrative Region (SAR) of China (Project
Nos. AoE/P701/20, C7037-22GF, 17302223, 17301924,
17301725), the ANR/RGC Joint Research Scheme sponsored by RGC of Hong Kong and French National Research Agency (Project No. A\_HKU703/22) and the State Key Laboratory of Optical Quantum Materials at HKU. S.S. was supported by the Simons Collaboration on Ultra-Quantum Matter which is a grant from the Simons Foundation (651440, S.S.). We thank
HPC2021 system under the Information Technology Services
at the University of Hong Kong~\cite{hpc2021}, as well as the Beijing Paratera Tech Corp., Ltd~\cite{paratera} for providing HPC resources that have contributed to the research results reported within this paper.

\bibliography{references}

\newpage


\section*{Supplementary material}

# Supplementary Information for

## Deconfined criticality between an antiferromagnetic insulator and a nodal d-wave superconductor: a quantum Monte Carlo study

**Chuang Chen, Subir Sachdev, Zi Yang Meng**

**Corresponding author: Subir Sachdev, E-mail: sachdev@g.harvard.edu**

### This PDF file includes:

Supplementary text
Figs. S1 to S3
References for SI reference citations

## Supporting Information Text

## 1. DQMC Formulation

In order to simulate the model of the main text using DQMC, we need to choose a representation for the SU(2) gauge field and to perform the trace over the fermions in the partition function $Z = \mathrm{Tr} e^{-\beta\mathcal{H}}$.

**A. Quaternion Representation.** There are two popular ways to parameterize SU(2) group elements: the quaternion parameterization, and the exponential map from the Lie algebra to the Lie group. We use the quaternion method, as it is better suited to Metropolis-type Monte Carlo updates, is easier to manipulate, and avoids repeated exponentiations. The exponential map is more appropriate for molecular-dynamics Hamiltonian Monte Carlo updates. We use four real numbers $(a_0, a_1, a_2, a_3)$ to represent an SU(2) element as $U = \begin{pmatrix} a_0 + ia_3 & ia_1 + a_2 \\ ia_1 - a_2 & a_0 - ia_3 \end{pmatrix} \in SU(2)$, subject to the constraint $a_0^2 + a_1^2 + a_2^2 + a_3^2 = 1$. A general SU(2) matrix $U_{ij}$ has the structure $\begin{pmatrix} u_{11} & u_{12} \\ -u_{12}^* & u_{11}^* \end{pmatrix}$, so that only three free parameters are needed to parameterize the SU(2) group elements.

**B. Basis Transformation.** We first expand the spinon–gauge coupling term $\mathcal{H}_f$,

$$\begin{aligned}\Psi_i^\dagger U_{ij}\Psi_j &= (f_{i\uparrow}^\dagger, f_{i\downarrow}) \begin{pmatrix} u_{11} & u_{12} \\ u_{21} & u_{22} \end{pmatrix} \begin{pmatrix} f_{j\uparrow} \\ f_{j\downarrow}^\dagger \end{pmatrix} \\ &= u_{11} f_{i\uparrow}^\dagger f_{j\uparrow} + u_{12} f_{i\uparrow}^\dagger f_{j\downarrow}^\dagger + u_{21} f_{i\downarrow} f_{j\uparrow} + u_{22} f_{i\downarrow} f_{j\downarrow}^\dagger\end{aligned}$$

where $u_{22} = u_{11}^*$ and $u_{21} = -u_{12}^*$. Because of the $f^\dagger f^\dagger$ and $ff$ terms in the expanded Hamiltonian, we cannot perform the trace directly. We are working at half filling, and can therefore apply a particle-hole transformation to the spin-down part, $f_{i\downarrow} \to (-1)^i \tilde{f}_{i\downarrow}^\dagger$, while leaving the spin-up part unchanged, apart from the redefinition $f_{i\uparrow} \to \tilde{f}_{i\uparrow}$. The $ff$ and $f^\dagger f^\dagger$ terms are then mapped onto $\tilde{f}^\dagger \tilde{f}$ as follows for a nearest-neighbor bond $\langle ij \rangle$:

$$\begin{aligned} f_{i\uparrow}^\dagger f_{j\uparrow} &\to \tilde{f}_{i\uparrow}^\dagger \tilde{f}_{j\uparrow} \\ f_{i\uparrow}^\dagger f_{j\downarrow}^\dagger &\to \tilde{f}_{i\uparrow}^\dagger (-1)^j \tilde{f}_{j\downarrow} \\ f_{i\downarrow} f_{j\uparrow} &\to (-1)^i \tilde{f}_{i\downarrow}^\dagger \tilde{f}_{j\uparrow} \\ f_{i\downarrow} f_{j\downarrow}^\dagger &\to (-1)^{i+j} \tilde{f}_{i\downarrow}^\dagger \tilde{f}_{j\downarrow} = -\tilde{f}_{i\downarrow}^\dagger \tilde{f}_{j\downarrow} \end{aligned}$$

Arranging all the fermion operators into a vector, we form the new basis $\vec{\tilde{f}}^\dagger = (\tilde{f}_{1\uparrow}^\dagger, \tilde{f}_{1\downarrow}^\dagger, \tilde{f}_{2\uparrow}^\dagger, \tilde{f}_{2\downarrow}^\dagger, \dots)$, in which $\mathcal{H}_{\tilde{f}} = \vec{\tilde{f}}^\dagger \mathcal{K}[U] \vec{\tilde{f}}$. The $2 \times 2$ blocks $K_{ij}$ of the full $2N \times 2N$ hopping matrix $\mathcal{K}$ on the nearest-neighbor bonds $\langle ij \rangle$ are

$$K_{ij} = iJe_{ij} \begin{pmatrix} u_{11} & (-1)^j u_{12} \\ (-1)^i u_{21} & (-1)^{i+j} u_{22} \end{pmatrix}, K_{ji} = -iJe_{ij} \begin{pmatrix} u_{11}^* & (-1)^i u_{21}^* \\ (-1)^j u_{12}^* & (-1)^{i+j} u_{22}^* \end{pmatrix} = K_{ij}^\dagger \qquad [S1]$$

All other blocks are $2 \times 2$ zero matrices, and $N = L^2$ is the number of lattice sites.

**C. Suzuki–Trotter Decomposition.** We now apply the Trotter decomposition and integrate out the fermions in the partition function,

$$\begin{aligned} Z &= \int \mathcal{D}[B]\mathcal{D}[U] \mathrm{Tr} e^{-\beta\mathcal{H}} \\ &= \int \mathcal{D}[B]\mathcal{D}[U] \mathrm{Tr} \prod_\tau e^{-\Delta\tau\mathcal{H}_B} e^{-\Delta\tau\mathcal{H}_U} e^{-\Delta\tau\mathcal{H}_f} + \mathcal{O}(\Delta\tau^2) \\ &= \det(1 + \prod_\tau e^{-\Delta\tau\mathcal{K}_\tau[U]}) \int \mathcal{D}[B]\mathcal{D}[U] \prod_\tau e^{-\Delta\tau\mathcal{H}_B} e^{-\Delta\tau\mathcal{H}_U} + \mathcal{O}(\Delta\tau^2) \end{aligned}$$

A sign-problem-free determinant QMC simulation requires det $\geq 0$ for every possible configuration of the SU(2) gauge field $U$. Acting on $\mathcal{H}_f$ with the antiunitary time-reversal transformation $\tilde{f}_{i\uparrow} \to \tilde{f}_{i\downarrow}, \tilde{f}_{i\downarrow} \to -\tilde{f}_{i\uparrow}, i \to -i$, we can confirm that $\mathcal{H}_f$ is invariant, which guarantees a sign-problem-free DQMC simulation in the $\vec{\tilde{f}}^\dagger = (\tilde{f}_{1\uparrow}^\dagger, \tilde{f}_{1\downarrow}^\dagger, \tilde{f}_{2\uparrow}^\dagger, \tilde{f}_{2\downarrow}^\dagger, \dots)$ basis. In order to use fast updates in the DQMC simulation, we further split the propagator $e^{-\Delta\tau\mathcal{K}_\tau}$ into a sparse representation. We sort the bonds of the square lattice into four families, namely up, down, left, right, each of which involves only half of the lattice sites, so that

$$e^{-\Delta\tau\mathcal{K}_\tau} = \prod_{\ell\in\mathcal{F}_1} e^{-\Delta\tau\mathbf{K}_{\tau,1,\ell}} \prod_{\ell\in\mathcal{F}_2} e^{-\Delta\tau\mathbf{K}_{\tau,2,\ell}} \prod_{\ell\in\mathcal{F}_3} e^{-\Delta\tau\mathbf{K}_{\tau,3,\ell}} \prod_{\ell\in\mathcal{F}_4} e^{-\Delta\tau\mathbf{K}_{\tau,4,\ell}} + \mathcal{O}(\Delta\tau^2).$$

Here $a = 1, 2, 3, 4$ labels a bond family and $\ell \in \mathcal{F}_a$ labels a bond within that family. Each factor $\mathbf{K}_{\tau,a,\ell}$ acts nontrivially only in a $4 \times 4$ active subspace and is block off-diagonal, with vanishing diagonal blocks. When it appears in the product above, $e^{-\Delta\tau \mathbf{K}_{\tau,a,\ell}}$ is understood to be embedded in the full one-particle space and to act as the identity on the complementary subspace. In order to evaluate the product $\prod_\tau e^{-\Delta\tau \mathcal{K}_\tau[U]}$, we need to diagonalize $\mathbf{K}_{\tau,a,\ell}$ in this active subspace, as follows. Recall that

$$K_{ij} = iJe_{ij} \begin{pmatrix} u_{11} & (-1)^j u_{12} \\ (-1)^i u_{21} & (-1)^{i+j} u_{22} \end{pmatrix} = iJe_{ij} \begin{pmatrix} u_{11} & (-1)^j u_{12} \\ -(-1)^j u_{21} & -u_{22} \end{pmatrix} = iJe_{ij} \begin{pmatrix} u_{11} & \epsilon_j u_{12} \\ \epsilon_j u_{12}^* & -u_{11}^* \end{pmatrix}, \tag{S2}$$

and

$$K_{ji} = -iJe_{ij} \begin{pmatrix} u_{11}^* & (-1)^i u_{21}^* \\ (-1)^j u_{12}^* & (-1)^{i+j} u_{22}^* \end{pmatrix} = K_{ij}^\dagger = -iJe_{ij} \begin{pmatrix} u_{11}^* & \epsilon_j u_{12} \\ \epsilon_j u_{12}^* & -u_{11} \end{pmatrix}. \tag{S3}$$

Thus, $K_{ji} = K_{ij}^\dagger$, which ensures that the full hopping matrix $\mathcal{K}$ is Hermitian. Here $\mathbf{K} \equiv \mathbf{K}_{\tau,a,\ell}$ is the $4 \times 4$ active block matrix whose diagonal blocks vanish and whose off-diagonal blocks are $K_{ij}$ and $K_{ji}$:

$$\mathbf{K} = \begin{pmatrix} 0 & 0 & iJe_{ij}\, u_{11} & iJe_{ij}\epsilon_j u_{12} \\ 0 & 0 & iJe_{ij}\epsilon_i u_{21} & iJe_{ij}\epsilon_i\epsilon_j u_{22} \\ -iJe_{ij}\, u_{11}^* & -iJe_{ij}\epsilon_i u_{21}^* & 0 & 0 \\ -iJe_{ij}\epsilon_j u_{12}^* & -iJe_{ij}\epsilon_i\epsilon_j u_{22}^* & 0 & 0 \end{pmatrix}. \tag{S4}$$

To evaluate $e^{-\Delta\tau\mathbf{K}}$, we first diagonalize $\mathbf{K}$ as

$$\mathbf{K} = UDU^{-1}, \tag{S5}$$

and we choose $U$ to be unitary, $U^{-1} = U^\dagger$, so that

$$e^{-\Delta\tau\mathbf{K}} = Ue^{-\Delta\tau D}U^\dagger. \tag{S6}$$

Writing $K_{12} \equiv K_{ij}$ for the upper-right $2 \times 2$ block of $\mathbf{K}$ for $\ell = \langle ij \rangle$, and noting that $\mathbf{K}$ is a block-off-diagonal Hermitian matrix, we have

$$\begin{gathered} \mathbf{K}v = \lambda v \\ \begin{pmatrix} 0 & K_{12} \\ K_{12}^\dagger & 0 \end{pmatrix} \begin{pmatrix} x \\ y \end{pmatrix} = \lambda \begin{pmatrix} x \\ y \end{pmatrix} \\ K_{12}^\dagger K_{12} y = \lambda^2 y \\ \lambda = \pm\sqrt{\text{eigvals}(K_{12}^\dagger K_{12})} \\ v = \begin{pmatrix} x \\ y \end{pmatrix} = \begin{pmatrix} \frac{1}{\lambda} K_{12} y \\ y \end{pmatrix} \end{gathered} \tag{S7}$$

The eigenvalues are

$$\lambda = \pm J \tag{S8}$$

i.e., two distinct eigenvalues, each twofold degenerate. The eigenvectors $y$ of $K_{12}^\dagger K_{12}$ are $(1, 0)$ and $(0, 1)$. After normalization and Gram–Schmidt orthogonalization, we obtain a unitary $U$. Explicitly, $D$ and $U$ are

$$\begin{aligned} D =& (+J, -J, +J, -J) \\ U =& \frac{1}{\sqrt{2}} \begin{pmatrix} ie_{ij}u_{11} & -ie_{ij}u_{11} & ie_{ij}\epsilon_j u_{12} & -ie_{ij}\epsilon_j u_{12} \\ ie_{ij}\epsilon_j u_{12}^* & -ie_{ij}\epsilon_j u_{12}^* & -ie_{ij}u_{11}^* & ie_{ij}u_{11}^* \\ 1 & 1 & 0 & 0 \\ 0 & 0 & 1 & 1 \end{pmatrix} \end{aligned}$$

**D. Observables.** The Green's function obtained from the DQMC simulation is expressed in the $\vec{\tilde{f}}$ basis, whereas the physical observables are defined in the original $\vec{f}$ basis. We therefore have to undo the particle-hole transformation and compute the observables from the original Green's functions. The mapping is as follows.

$$\begin{aligned} \left\langle \tilde{f}_{i\uparrow}\tilde{f}_{j\uparrow}^\dagger \right\rangle &\Longleftrightarrow \left\langle f_{i\uparrow}f_{j\uparrow}^\dagger \right\rangle, \\ \left\langle \tilde{f}_{i\uparrow}\tilde{f}_{j\downarrow}^\dagger \right\rangle &\Longleftrightarrow (-1)^j \left\langle f_{i\uparrow}f_{j\downarrow} \right\rangle, \\ \left\langle \tilde{f}_{i\downarrow}\tilde{f}_{j\uparrow}^\dagger \right\rangle &\Longleftrightarrow (-1)^i \left\langle f_{i\downarrow}^\dagger f_{j\uparrow}^\dagger \right\rangle, \\ \left\langle \tilde{f}_{i\downarrow}\tilde{f}_{j\downarrow}^\dagger \right\rangle &\Longleftrightarrow (-1)^{i+j} \left\langle f_{i\downarrow}^\dagger f_{j\downarrow} \right\rangle. \end{aligned} \tag{S9}$$

This mapping is needed for both equal-time and dynamical measurements. In the simulation we measure the spin correlation function $S_i^+ S_j^-$, and the dimer correlation function $D_i D_j$, and obtain the corresponding structure factors. We use Wick's

theorem to decompose the multi-fermion correlators into products of fermion-bilinear Green's functions. Usually one need only consider contractions that generate normal Green's functions, such as $G(i,j) = \langle f_i f_j^\dagger \rangle$ and $\bar{G}(i,j) = \langle f_i^\dagger f_j \rangle$; in our model, however, contractions corresponding to the anomalous Green's functions $F(i,j) = \langle f_i f_j \rangle$ and $\bar{F}(i,j) = \langle f_i^\dagger f_j^\dagger \rangle$ are unavoidable. As a consequence, the Wick-decomposed expression for a correlation function contains more terms. Take the spin correlation as an example:

$$\begin{aligned} \langle S_i^+ S_j^- \rangle &= \langle f_{i,\uparrow}^\dagger f_{i,\downarrow} f_{j,\downarrow}^\dagger f_{j,\uparrow} \rangle \\ &= -F_{\downarrow\uparrow}(i,j) \cdot \bar{F}_{\uparrow\downarrow}(i,j) + G_\downarrow(i,j) \cdot \bar{G}_\uparrow(i,j) \end{aligned} \quad \text{[S10]}$$

The dimer correlation $\langle D_i D_j \rangle = \langle (S_i^x S_{i+\hat{x}}^x + S_i^y S_{i+\hat{x}}^y + S_i^z S_{i+\hat{x}}^z) \cdot (S_j^x S_{j+\hat{x}}^x + S_j^y S_{j+\hat{x}}^y + S_j^z S_{j+\hat{x}}^z) \rangle$ contains many more terms, and therefore requires an automated program to perform the Wick decomposition.

## 2. Update Methods in DQMC

We need to update both the SU(2) gauge field $U$ and the chargon field $B$. Since $B$ does not couple to the fermions directly, its update is cheaper than that of $U$. The SU(2) gauge field $U$ couples to the fermions and enters the determinant in the partition function. Computing the acceptance ratio for a $U$ update therefore requires more intensive computation and, more importantly, once a proposal is accepted the equal-time Green's function has to be updated by a matrix-matrix multiplication. The overall computational complexity is $\mathcal{O}(\beta N^3)$, which makes large system sizes much harder to reach.

**A. SU(2) gauge field.** The gauge field couples to the fermions and enters the determinant, so its update inevitably involves computing a determinant ratio and refreshing the Green's function. Our scheme is to combine traditional Metropolis fast updates with global updates, so as to explore the gauge-field configurations efficiently while keeping the computational cost under control.

***A.1. Random-walk Metropolis fast update.*** Within the quaternion parameterization, an SU(2) group element can be generated by drawing $(a_0, a_1, a_2, a_3)$ from a Gaussian distribution and normalizing it by $\frac{1}{\sqrt{a_0^2+a_1^2+a_2^2+a_3^2}}$. For a local update, we instead generate a small SU(2) rotation from the Lie algebra: we draw $\epsilon_\alpha$ ($\alpha = 1, 2, 3$) from a Gaussian distribution with width $\delta_R$, set $\theta = |\boldsymbol{\epsilon}|$ and $\hat{\boldsymbol{n}} = \boldsymbol{\epsilon}/\theta$, and construct $R = \cos\theta\, I + i \sin\theta\, \hat{\boldsymbol{n}} \cdot \boldsymbol{\sigma}$ (with $R = I$ for $\theta = 0$). The update is then

$$U'_{ij} = RU_{ij} = (1 + R - 1)U_{ij} \quad \text{[S11]}$$

where we choose left multiplication throughout the simulation. Denote by $\mathcal{B}(\tau_2, \tau_1)$ the time-ordered fermion propagator from $\tau_1$ to $\tau_2$. We now work out the $\Delta$ matrix of the DQMC,

$$\det[1 + \mathcal{B}(\beta,\tau)\mathcal{B}(\tau,0)] \to \det[1 + \mathcal{B}(\beta,\tau)(1+\Delta)\mathcal{B}(\tau,0)]. \quad \text{[S12]}$$

Because of the way we split the propagator, a local Metropolis update of the gauge link $U_{ij}$ affects only a single $4 \times 4$ active matrix $\mathbf{K}_{\tau,a,\ell}$, with $\ell = \langle ij \rangle$, so the effective $\Delta$ matrix is also only $4 \times 4$. We can make use of the special block form of the off-diagonal Hermitian matrix,

$$e^{-\Delta\tau\mathbf{K}} = \begin{pmatrix} \cosh(\Delta\tau\sqrt{K_{12}K_{12}^\dagger}) & -\frac{\sinh(\Delta\tau\sqrt{K_{12}K_{12}^\dagger})}{\sqrt{K_{12}K_{12}^\dagger}} K_{12} \\ -K_{12}^\dagger \frac{\sinh(\Delta\tau\sqrt{K_{12}K_{12}^\dagger})}{\sqrt{K_{12}K_{12}^\dagger}} & \cosh(\Delta\tau\sqrt{K_{12}^\dagger K_{12}}) \end{pmatrix}, K_{12}K_{12}^\dagger = J^2 I_2, K_{12}^\dagger K_{12} = J^2 I_2 \quad \text{[S13]}$$

$$= \begin{pmatrix} \cosh(\Delta\tau J) I_2 & -\frac{\sinh(\Delta\tau J)}{J} I_2 K_{12} \\ -K_{12}^\dagger \frac{\sinh(\Delta\tau J)}{J} I_2 & \cosh(\Delta\tau J) I_2 \end{pmatrix} \quad \text{[S14]}$$

We now determine the $\Delta$ matrix,

$$\Delta = e^{-\Delta_\tau \mathbf{K}'} e^{\Delta_\tau \mathbf{K}} - I_4$$

$$\begin{pmatrix} \Delta_{11} & \Delta_{12} \\ \Delta_{21} & \Delta_{22} \end{pmatrix} = \begin{pmatrix} \cosh(\Delta\tau J) I_2 & -\frac{\sinh(\Delta\tau J)}{J} I_2 K'_{12} \\ -K'^\dagger_{12} \frac{\sinh(\Delta\tau J)}{J} I_2 & \cosh(\Delta\tau J) I_2 \end{pmatrix} \cdot \begin{pmatrix} \cosh(\Delta\tau J) I_2 & \frac{\sinh(\Delta\tau J)}{J} I_2 K_{12} \\ K_{12}^\dagger \frac{\sinh(\Delta\tau J)}{J} I_2 & \cosh(\Delta\tau J) I_2 \end{pmatrix} - \begin{pmatrix} I_2 & 0 \\ 0 & I_2 \end{pmatrix}$$

with the result

$$\Delta_{11} = \sinh^2(\Delta\tau J) I_2 - \frac{\sinh^2(\Delta\tau J)}{J^2} K'_{12} K_{12}^\dagger$$

$$\Delta_{12} = \cosh(\Delta\tau J) \frac{\sinh(\Delta\tau J)}{J} (K_{12} - K'_{12})$$

$$\Delta_{21} = \cosh(\Delta\tau J) \frac{\sinh(\Delta\tau J)}{J} (K_{12}^\dagger - K'^\dagger_{12})$$

$$\Delta_{22} = \sinh^2(\Delta\tau J) I_2 - \frac{\sinh^2(\Delta\tau J)}{J^2} K'^\dagger_{12} K_{12}$$

We rewrite $K_{12}$ as

$$K_{12} = iJe_{ij}\begin{pmatrix} u_{11} & \epsilon_j u_{12} \\ \epsilon_j u_{12}^* & -u_{11}^* \end{pmatrix} = iJe_{ij}\begin{pmatrix} 1 & 0 \\ 0 & -\epsilon_j \end{pmatrix}\begin{pmatrix} u_{11} & u_{12} \\ -u_{12}^* & u_{11}^* \end{pmatrix}\begin{pmatrix} 1 & 0 \\ 0 & \epsilon_j \end{pmatrix} = iJe_{ij}\begin{pmatrix} 1 & 0 \\ 0 & -\epsilon_j \end{pmatrix} M \begin{pmatrix} 1 & 0 \\ 0 & \epsilon_j \end{pmatrix} = iJe_{ij}D_1 M D_2 \tag{S15}$$

With the Metropolis update $M' = RM$, and noting that $[R, M] \neq 0$ since SU(2) is a non-Abelian Lie group, we obtain

$$\begin{aligned} K'_{12}K^\dagger_{12} &= J^2 D_1 R M D_2 D_2^\dagger M^\dagger D_1^\dagger \\ &= J^2 D_1 R D_1^\dagger \\ &= J^2 \begin{pmatrix} r_{11} & -\epsilon_j r_{12} \\ \epsilon_j r_{12}^* & r_{11}^* \end{pmatrix} \\ K'^\dagger_{12} K_{12} &= J^2 D_2^\dagger M^\dagger R^\dagger D_1^\dagger D_1 M D_2 \\ &= J^2 D_2^\dagger M^\dagger R^\dagger M D_2 \end{aligned}$$

$$\begin{aligned} K_{12} - K'_{12} &= iJe_{ij}D_1 M D_2 - iJe_{ij}D_1 R M D_2 \\ &= iJe_{ij}D_1(M - RM)D_2 \\ &= iJe_{ij}D_1(I - R)MD_2 \end{aligned}$$

$$\begin{aligned} K^\dagger_{12} - K'^\dagger_{12} &= -iJe_{ij}D_2 M^\dagger D_1 - (-iJe_{ij}D_2 M^\dagger R^\dagger D_1) \\ &= -iJe_{ij}D_2\left(M^\dagger - M^\dagger R^\dagger\right)D_1 \\ &= -iJe_{ij}D_2 M^\dagger (I - R^\dagger) D_1 \end{aligned}$$

Note that $K'_{12}K^\dagger_{12}$ is independent of $M$. For $K'^\dagger_{12}K_{12}$, one has to perform a matrix multiplication. The other two terms involve only matrix subtractions and are easy to evaluate, while $M' = RM$ again requires a matrix multiplication. Once the $4 \times 4$ $\Delta$ matrix is known, the calculation of the determinant ratio and the updating of the equal-time Green's function are routine.

***A.2. Over-relaxation Metropolis fast update.*** For small $\kappa_\tau$, a large random-walk update is very likely to be rejected. It is therefore worthwhile to design an update method that leaves $\mathcal{H}_{U_\tau}$ unchanged while still allowing large moves of the gauge field. The overrelaxation update has exactly this property within a Metropolis scheme and substantially improves the sampling at small $\kappa_\tau$, as in the data shown in Fig. 2 of the main text. We adopt the following update,

$$U \to U' = V^\dagger U^\dagger V^\dagger \tag{S16}$$

The temporal Wilson action is $\mathrm{Tr}[U(\tau)U^\dagger(\tau+1) + U(\tau)U^\dagger(\tau-1)] = \mathrm{Tr}[UA]$ with $A = U^\dagger(\tau+1) + U^\dagger(\tau-1)$. We construct $V = A/\sqrt{\det[A]}$, which is unitary, i.e., $V^{-1} = V^\dagger$. We then have

$$\mathrm{Tr}[U'A] = \mathrm{Tr}[V^\dagger U^\dagger V^\dagger A] = \sqrt{\det[A]}\mathrm{Tr}[V^\dagger U^\dagger] = \mathrm{Tr}[A^\dagger U^\dagger] = \mathrm{Tr}[UA] \tag{S17}$$

In order to use the same framework as in the random-walk Metropolis update, we can set $R = V^\dagger U^\dagger V^\dagger U^\dagger$ and proceed to construct the $\Delta$ matrix. Note that the traces of SU(2) matrices are real. With the overrelaxation update the temporal Wilson action is unchanged. The spatial gauge and chargon contributions to the action, together with the fermion determinant, are included in the Metropolis acceptance probability. Overrelaxation alone is not ergodic, however, and must be combined with ordinary updates.

***A.3. Block global update.*** Beyond the two local Metropolis updates described above, at small $\kappa_\tau$ we can update the gauge field efficiently by means of a so-called global update, which updates a whole block, or chain, of gauge fields at once. Specifically, we fix one lattice bond $ij$ and update the gauge field $U_{ij,\tau}$ on all time slices with the same matrix $R$,

$$U_{ij,\tau} \to U'_{ij,\tau} = RU_{ij,\tau} \tag{S18}$$

The block update also leaves $\mathcal{H}_{U_\tau}$ unchanged. The determinant ratio, however, can no longer be computed as in the Metropolis fast update: one has to evaluate the determinants themselves and then take their ratio, at a computational cost of $\mathcal{O}(N^3)$. The block update is thus expensive, but it does make large moves of the gauge field possible.

**B. Chargon.** The chargon does not couple to the fermions directly and does not enter the determinant. In principle, updating the chargon field is therefore computationally cheaper and permits more tricks. Here too we combine local Metropolis updates with global updates (which are much simpler than in the SU(2) case) in order to update the chargon field efficiently.

***B.1. Random-walk Metropolis update.*** The chargon field $B$ has two components $B_i$, each of which is a complex number. For $r < 0$, the norm of the chargon field, $\langle |B| \rangle$, is finite, and the different ordered phases correspond to different patterns of the relative phase between the chargon and the SU(2) gauge field. In order to update the chargon field effectively, we treat its norm and its internal SU(2) orientation separately. We first update the norm,

$$B_i \to B_i' = (1 + \eta\delta_B)\, B_i \qquad [S19]$$

where $\eta$ is a standard Gaussian random number and $\delta_B$ controls the update size. The internal orientation is then updated by rotating the chargon vector $B$ with an SU(2) rotation,

$$B_i \to B_i' = R B_i \qquad [S20]$$

where $R$ is a random SU(2) matrix. This operation preserves the chargon norm and changes only its internal orientation.

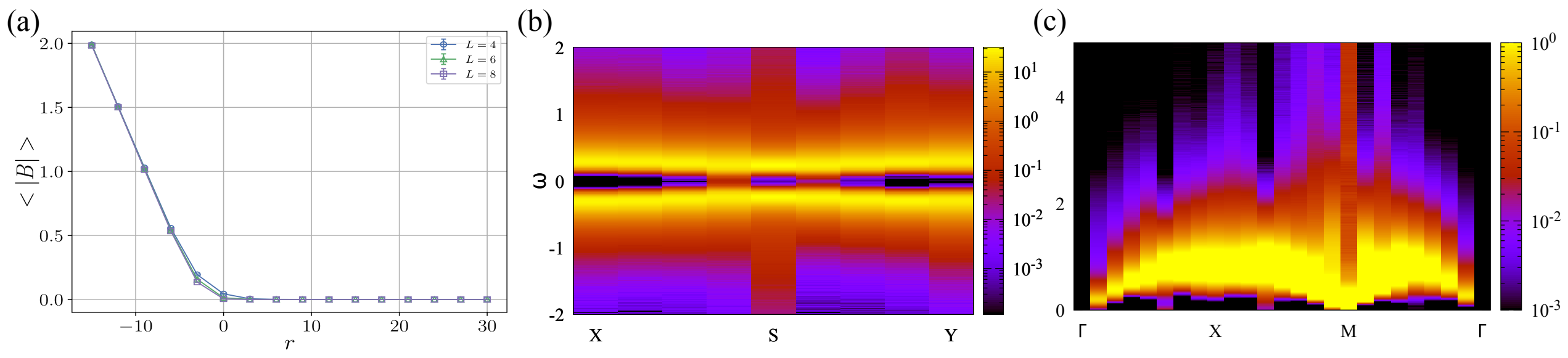


**Fig. S1. Chargon amplitude and spectra in the AFM phase at negative $r$.** (a) $\langle |B| \rangle$ as a function of $r$ at $\kappa_\tau = 10$. As $r$ is increased from negative to positive values, $\langle |B| \rangle$ decreases continuously from a finite value to nearly zero. (b) Electron spectrum $A(\mathbf{k}, \omega)$ in the AFM phase at $r = -15$, $\kappa_\tau = 10$, $L = 16$, and $\beta = 24$. (c) Corresponding spin spectrum.

***B.2. Block global update.*** For small $m$ it becomes harder to update the chargon field, because of the time-derivative term $\mathcal{H}_{B_\tau}$, in analogy with the $\mathcal{H}_{U_\tau}$ case. Here too we can design a global block update that leaves $\mathcal{H}_{B_\tau}$ unchanged, and again it is necessary to update the norm and the internal orientation separately. For the orientation update we can proceed as in the local Metropolis update, updating the chargon fields that belong to the same lattice site across all time slices,

$$B_{i,\tau} \to B_{i,\tau}' = R B_{i,\tau} \qquad [S21]$$

Updating the norm on all time slices in this way, however, appears to be impossible, because of the periodic boundary condition in the imaginary-time direction. The best we can do is to update the norm by a translation in vector space, which at the same time also changes the internal orientation,

$$B_{i,\tau} \to B_{i,\tau}' = B_{i,\tau} + b \qquad [S22]$$

where $b$ is a random vector generated by a Gaussian random number generator. We find that the global update is essential for obtaining results in the quantum case that are consistent with those of the classical model.

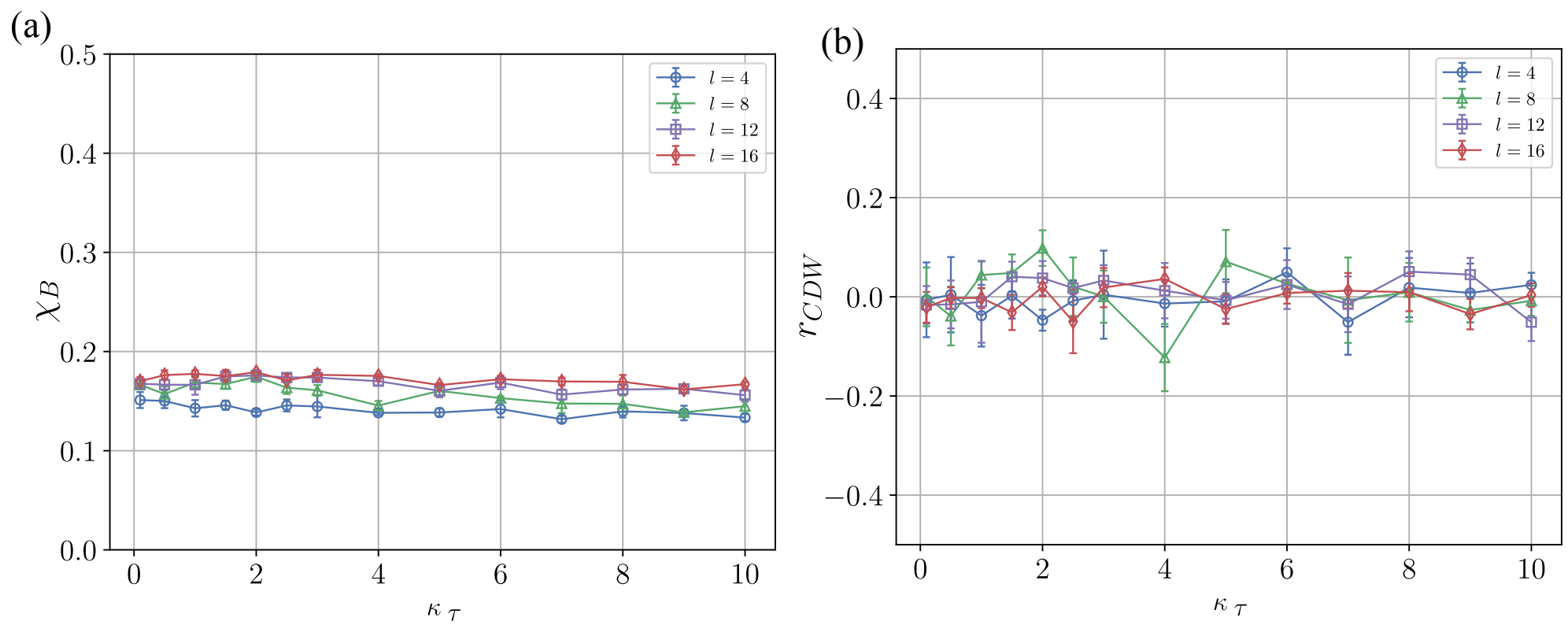


**Fig. S2. CDW structure factor and correlation ratio across the dSC-to-AFM transition.** (a) CDW structure factor $\chi_B$ at the ordering wave vectors $\mathbf{q} = \mathbf{X}$ and $\mathbf{Y}$ as a function of $\kappa_\tau$. The absence of systematic growth with system size indicates that CDW correlations do not become long ranged. (b) Corresponding CDW correlation ratio $r_{\mathrm{CDW}}$. It remains close to zero throughout the range of $\kappa_\tau$, providing no evidence for enhanced CDW order at either $\mathbf{X}$ or $\mathbf{Y}$.

## 3. More Numerical Data

**A. Spectra at $r = -15, \kappa_\tau = 10$.** We note that, at large $\kappa_\tau$, the AFM phase in the phase diagram extends over both positive and negative values of $r$. In the main text, we presented the electron and spin spectra for positive $r$, which display the characteristic features of the AFM phase. Here we provide complementary spectral data at negative $r$. The amplitude of the chargon field, $\langle|B|\rangle$, is controlled by $r$: it is finite for negative $r$ and approaches zero for positive $r$. Since the chargon field enters directly into the electron Green's function, and hence the spectral function $A(\mathbf{k}, \omega)$, the electron spectral weight is sensitive to the value of $r$. Fig. S1 shows $\langle|B|\rangle$ as a function of $r$ at $\kappa_\tau = 10$, together with the electron and spin spectra at $r = -15$, $\kappa_\tau = 10$, $L = 16$, and $\beta = 24$. For negative $r$, $\langle|B|\rangle$ is clearly finite, and the electron spectrum carries much larger spectral weight than in the positive-$r$ case shown in the main text. By contrast, the spin spectrum is only weakly affected by changing $r$. Thus, provided that $\kappa_\tau$ is sufficiently large, the system remains in the AFM phase for both signs of $r$, with qualitatively the same electron and spin dynamical properties.

**B. Charge density wave.** We examine charge-density-wave (CDW) correlations across the dSC-to-AFM transition by varying $\kappa_\tau$. We define the CDW structure factor as

$$\chi_B(\mathbf{q}) = \frac{1}{L^2}\sum_{i,j}\langle\rho_i\rho_j\rangle e^{-i\mathbf{q}\cdot(\mathbf{r}_i-\mathbf{r}_j)}, \qquad \rho_i = B_i^\dagger B_i, \tag{S23}$$

and evaluate it at the candidate ordering wave vectors $\mathbf{q} = \mathbf{X} = (\pi, 0)$ and $\mathbf{q} = \mathbf{Y} = (0, \pi)$. As shown in Fig. S2(a), $\chi_B$ remains nearly independent of $\kappa_\tau$ and exhibits no systematic growth with system size. The corresponding correlation ratio $r_{\mathrm{CDW}}$ also fluctuates around zero throughout the transition [Fig. S2(b)]. These results provide no evidence for CDW ordering or for an enhancement of CDW correlations near the dSC-to-AFM transition.

**C. Binder cumulant of the dSC order.** To provide an additional diagnostic of the dSC-to-AFM transition, we measure the Binder cumulant of the dSC order parameter. Using the local d-wave pairing field defined in the main text, the quantity sampled in each QMC measurement is its spatial average

$$\mathcal{D} = \frac{1}{L^2}\sum_i \Delta_i. \tag{S24}$$

The Binder cumulant is then defined as

$$U_{\mathrm{el,d}} = 1 - \frac{\langle|\mathcal{D}|^4\rangle}{2\langle|\mathcal{D}|^2\rangle^2}. \tag{S25}$$

In the QMC implementation, $|\mathcal{D}|^2$ and $|\mathcal{D}|^4$ are accumulated for each measurement and averaged independently before evaluating Eq. (S25); this is the quantity plotted in Fig. S3. For a complex dSC order parameter, $U_{\mathrm{el,d}}$ approaches 1/2 when the dSC order is well developed, whereas it approaches zero in the AFM phase, where the dSC order is absent. The Binder cumulant decreases from values close to 1/2 on the dSC side to values near zero on the AFM side as $\kappa_\tau$ is increased. Importantly, the curves do not show the negative dip expected for a clear first-order transition. Thus, these data provide additional evidence consistent with a continuous dSC-to-AFM transition. However, given the accessible system sizes, the Binder data alone cannot unambiguously distinguish a continuous transition from weak first-order behavior or a narrow coexistence region.

## 4. Quantum field theory of the DQCP

With the aim of aiding possible future numerical studies using bootstrap or fuzzy sphere methods, here we review here the structure of the proposed CFT, $\mathcal{L}_{\mathcal{X}} + \mathcal{L}_{\mathcal{B}}$ below, describing the DQCP.

We describe the fermionic spinons using a notation closely related to that of Section 3.2.1 in Ref. (1). The fermionic spinons are represented by the Dirac fermions $\mathcal{X}_v$, where $v$ is a double-valued valley/sublattice index chosen to span a space which is tensor product to the space of the Dirac matrices, $\gamma^\mu$. Each fermion $\mathcal{X}_v$ is a $2\times 2$ matrix in the SU(2) gauge/spin space. We represent Pauli matrices in valley space by $\mu^\ell$, and in the SU(2) gauge/spin space by $\sigma^\ell$, and the $\mu^\ell$, $\sigma^\ell$, and $\gamma^\mu$ all commute with each other. A crucial constraint on $\mathcal{X}_v$ is the reality condition

$$\mathcal{X}_v^\dagger = \sigma^y \mathcal{X}_v^T \sigma^y\,. \tag{S26}$$

The continuum theory for the Dirac fermionic spinons is

$$\mathcal{L}_{\mathcal{X}} = \frac{i}{2}\mathrm{Tr}\left(\bar{\mathcal{X}}_v\gamma^\mu\left[\partial_\mu - iA_\mu^\ell\sigma^\ell\right]\mathcal{X}_v\right)\,, \tag{S27}$$

where $A_\mu^\ell$ is the SU(2) gauge field. Therefore, the SU(2) gauge transformation acts on $\mathcal{X}$ by left multiplication

$$\mathcal{X}_v \to \exp(i\theta_\ell\sigma^\ell)\mathcal{X}_v\,. \tag{S28}$$

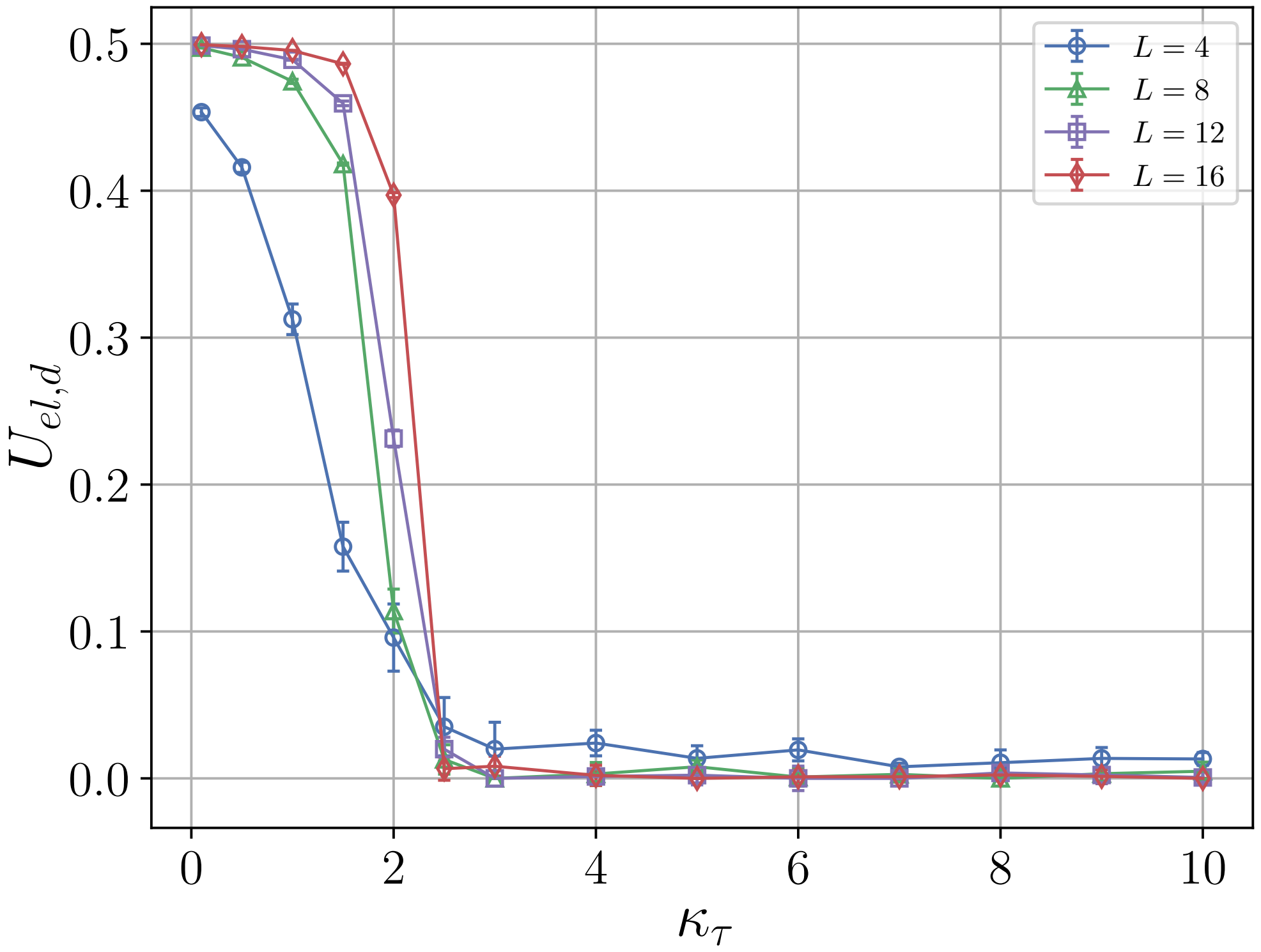


**Fig. S3. Binder cumulant of the dSC order across the dSC-to-AFM transition.** The Binder cumulant $U_{\mathrm{el,d}}$ is computed from the spatially averaged dSC order parameter according to $U_{\mathrm{el,d}} = 1 - \langle|\mathcal{D}|^4\rangle/[2\langle|\mathcal{D}|^2\rangle^2]$. The data are obtained along the same parameter path as the dSC-to-AFM scan in the main text, for system sizes $L = 4, 8, 12, 16$. The absence of a negative Binder-cumulant dip provides additional evidence consistent with a continuous transition, while not conclusively excluding weak first-order behavior or finite-size crossover effects.

There is a global USp(4) symmetry which acts on $\mathcal{X}$ by right multiplication in both the $\sigma^\ell$ and $\mu^\ell$ spaces:

$$\mathcal{X} \to \mathcal{X} \exp(i\vartheta_m M^m)\,. \tag{S29}$$

The Hermitian matrices $M^m$ should preserve the relation in Eq. (S26), leading to the constraint

$$M^{mT} = -\sigma^y M^m \sigma^y\,; \tag{S30}$$

this yields the 10 generators of USp(4)

$$M^m = \{\sigma^x, \sigma^y, \sigma^z, \mu^x\sigma^x, \mu^x\sigma^y, \mu^x\sigma^z, \mu^z\sigma^x, \mu^z\sigma^y, \mu^z\sigma^z, \mu^y\}\,. \tag{S31}$$

The global spin rotation symmetry of the underlying electronic model is a subgroup of this USp(4).

There is a parallel structure in the bosonic chargon sector. In the body of the paper, we introduced 4 complex scalars $B_{as}$ where $a$ is the fundamental SU(2) gauge index, and $s$ is a separate valley index taking 2 values. As for the fermions it is useful to write them in matrix form

$$\mathcal{B}_s \equiv \begin{pmatrix} B_{1s} & -B_{2s}^* \\ B_{2s} & B_{1s}^* \end{pmatrix}\,, \tag{S32}$$

so that we again have a reality constraint analogous to Eq. (S26)

$$\mathcal{B}_s^\dagger = \sigma^y \mathcal{B}_s^T \sigma^y\,. \tag{S33}$$

We introduce Pauli matrices $\rho^\ell$ to act on the $s$ valley space. Then the SU(2) gauge transformation is similar to Eq. (S28)

$$\mathcal{B}_s \to \exp(i\theta_\ell\sigma^\ell)\mathcal{B}_s\,, \tag{S34}$$

while the global symmetry transformations are similar to Eq. (S29)

$$\mathcal{B} \to \mathcal{B}\exp(i\vartheta_m M_c^m)\,. \tag{S35}$$

We can now define an algebra of $\mathrm{USp}(4)_c$ operators analogous to Eq. (S31)

$$M_c^m = \{\sigma^x, \sigma^y, \sigma^z, \rho^x\sigma^x, \rho^x\sigma^y, \rho^x\sigma^z, \rho^z\sigma^x, \rho^z\sigma^y, \rho^z\sigma^z, \rho^y\}\,. \tag{S36}$$

The chargon Lagrangian in the main text

$$\begin{aligned}\mathcal{L}_B =& |(\partial_\mu - iA^\alpha_\mu \sigma^\alpha)B|^2 + \bar{r}|B|^2 + \bar{u}\left(|B|^2\right)^2 \\ &+ \bar{v}_1 \left|\varepsilon_{ab}\varepsilon_{st} B_{as} B_{bt}\right|^2 \\ &+ \bar{v}_2 \left[(B^*_{a1}B_{a1} - B^*_{a2}B_{a2})^2 + (B^*_{a1}B_{a2} + B^*_{a2}B_{a1})^2\right] ,\end{aligned} \quad \text{[S37]}$$

has the full USp$(4)_c$ symmetry only for $\bar{v}_{1,2} = 0$.

If we have $\bar{v}_1 \neq 0$ and $\bar{v}_2 = 0$, then the USp$(4)_c$ symmetry is broken down to U$(1)_c\times$SU$(2)_c$. The U$(1)_c$ symmetry is associated with the charge conservation in the underlying electronic model under which $B \to e^{i\vartheta}B$. The SU$(2)_c$ symmetry, $B \to \exp(i\vartheta_\ell \rho^\ell)B$, rotates between the CDW$_{x,y}$ and staggered current order parameters which will be recalled below. In terms of the operators on the matrix field $\mathcal{B}$ in Eq. (S35), the symmetry is generated by a subalgebra of Eq. (S36)

$$\bar{M}^m_c = \{\sigma^z, \rho^x\sigma^z, \rho^z\sigma^z, \rho^y\} . \quad \text{[S38]}$$

If we have both $\bar{v}_1 \neq 0$ and $\bar{v}_2 \neq 0$, then the U$(1)_c$ symmetry is preserved, but the SU$(2)_c$ symmetry is broken down to the Pin(2) symmetry generated by $\exp(i\vartheta\rho^y)$ and $\exp(i(\pi/2)\rho^x\sigma^z)$.

To summarize, the largest global symmetry of the theory is USp(4)$\times$USp$(4)_c$.

- The USp(4) symmetry contains the SU(2) spin rotation symmetry of the underlying electrons as a subgroup. This USp(4) symmetry is anomalous, as it acts on Dirac fermions, and leads to a Wess-Zumino-Witten term when the fermions are integrated out.
- The USp$(4)_c$ symmetry contains the U$(1)_c$ charge conservation symmetry of the underlying electrons as a subgroup. This USp$(4)_c$ is not anomalous, as it acts on bosons. In general, the USp$(4)_c$ symmetry will be broken down to smaller subgroups, as noted above.

**Operator content of the CFT.** We can now present the simplest local, gauge-invariant operators, information which is needed for bootstrap or fuzzy sphere analysis of the CFT.

- There is a 5-component real scalar field $\Phi^k$, which transforms as a **5** of USp(4) *i.e.* as a vector under SO(5):

  $$\Phi^k = i\mathrm{Tr}\left(\mathcal{X}\Gamma^k\bar{\mathcal{X}}\right) \quad , \quad \Gamma^k = \{\mu^x, \mu^z, \mu^y\sigma^x, \mu^y\sigma^y, \mu^y\sigma^z\} \quad \text{[S39]}$$

  Here $\Phi^{3,4,5}$ represent the Néel order, while $\Phi^{1,2}$ are the valence bond solid orders. As noted above, the USp(4) symmetry is anomalous.
- There is a 5-component real scalar field $\Phi^k_c$, which transforms as a **5** of USp$(4)_c$ *i.e.* as a vector under SO$(5)_c$:

  $$\Phi^k_c = \mathrm{Tr}\left(\mathcal{B}\Gamma^k_c\mathcal{B}^\dagger\right) \quad , \quad \Gamma^k_c = \{\rho^x, \rho^z, \rho^y\sigma^x, \rho^y\sigma^y, \rho^y\sigma^z\} \quad \text{[S40]}$$

  The $\Phi^{1,2}_c$ form the CDW order parameters in the $x$ and $y$ directions, $B^*_{a1}B_{a1} - B^*_{a2}B_{a2}$ and $B^*_{a1}B_{a2} + B^*_{a2}B_{a1}$. The $\Phi^{3,4}_c$ are the complex dSC order parameter $\varepsilon_{ab}\varepsilon_{st}B_{as}B_{bt}$. And finally, $\Phi^5_c$ is the staggered current order, $i(B^*_{a1}B_{a2} - B^*_{a2}B_{a1})$. The dSC order parameter carries charge 2 under the U$(1)_c$ subgroup of USp$(4)_c$ associated with $\sigma^z$ in Eq. (S38), while the other scalars are neutral. As noted above, the USp$(4)_c$ symmetry is not anomalous.
- The novel feature of the Néel-dSC DQCP, not found in the usual Néel-VBS DQCP, is that there is also a gauge-invariant fermionic operator. This is a 16-component relativistic Majorana field, which transforms as a $\mathbf{4}\times\mathbf{4}$ of USp(4)$\times$USp$(4)_c$. This fermion carries charge 1 under the U$(1)_c$ subgroup of USp$(4)_c$ associated with $\sigma^z$ in Eq. (S38). This field is obtained from the real and imaginary components of the gauge-invariant matrix Dirac fermion field

  $$\mathcal{M}_{sv} = \mathcal{B}^\dagger_s \mathcal{X}_v \quad \text{[S41]}$$

  which also obeys the analog of Eq. (S26)

  $$\mathcal{M}^\dagger_{sv} = \sigma^y \mathcal{M}^T_{sv} \sigma^y . \quad \text{[S42]}$$

The Néel-VBS order parameter $\Phi^k$ can be obtained from the fusion of $\mathcal{M}$ and $\bar{\mathcal{M}}$ because integrating out Dirac fermions leads to an anomalous vector USp(4) symmetry. But conversely, the charge orders $\Phi^k_c$ are not in the fusion of $\mathcal{M}$ and $\bar{\mathcal{M}}$ because the charge orders are not anomalous.



\end{document}